\documentclass{article}
\pdfoutput=1
\pdfpageattr{/Group << /S /Transparency /I true /CS /DeviceRGB>>}
\usepackage{amssymb,amsmath}
\usepackage{graphicx}
\usepackage{amsfonts}
\usepackage{latexsym}
\usepackage{amsthm}
\usepackage{url}
\usepackage[iso]{isodateo}
\usepackage{capt-of}
\usepackage{hyperref}
\usepackage{epstopdf}
\usepackage{multirow}
\usepackage{tabularx}
\usepackage{algorithm}
\usepackage{algorithmicx}
\usepackage{algpseudocode}
\providecommand{\keywords}[1]{\textbf{\textit{keywords:}} #1}
\newcolumntype{C}{>{\centering\arraybackslash}p{2em}}

\title{A three domain covariance framework for EEG/MEG data}
\author{Beata Ro\'s\textsuperscript{a}, Fetsje Bijma\textsuperscript{a}, Mathisca C.M. de Gunst\textsuperscript{a}\thanks{Both authors contributed equally} \text{ } and Jan C. de Munck\textsuperscript{b}\footnotemark[1] }

\begin{document}
\maketitle
\noindent\textsuperscript{a} Department of Mathematics, Faculty of Exact Sciences, VU University Amsterdam, De Boelelaan 1081, 1081 HV, Amsterdam, The Netherlands\\
\textsuperscript{b} Department of Physics and Medical Technology, VU University Medical Center, De Boelelaan 1118,1081 HZ Amsterdam, The Netherlands\\
%\hrulefill
\begin{abstract}
In this paper we introduce a covariance framework for the analysis of EEG and MEG data that takes into account observed temporal stationarity on small time scales and trial-to-trial variations. We formulate a model for the covariance matrix, which is a Kronecker product of three components that correspond to space, time and epochs/trials, and consider maximum likelihood estimation of the unknown parameter values. An iterative algorithm that finds approximations of the maximum likelihood estimates is proposed. We perform a simulation study to assess the performance of the estimator and investigate the influence of different assumptions about the covariance factors on the estimated covariance matrix and on its components. Apart from that, we illustrate our method on real EEG and MEG data sets.

The proposed covariance model is applicable in a variety of cases where spontaneous EEG or MEG acts as source of noise and realistic noise covariance estimates are needed for accurate dipole localization, such as in evoked activity studies, or where the properties of spontaneous EEG or MEG are themselves the topic of interest, such as in combined EEG/fMRI experiments in which the correlation between EEG and fMRI signals is investigated.\\\\
\end{abstract}
%\hrulefill
\section{Introduction}
\noindent
In this work we propose a model for the covariance matrix of EEG/MEG that expresses it as a Kronecker product of three components and imposes physiologically inspired constraints on the model components. 
The analysis of the spatial distributions of EEG or MEG signals using mathematical models can provide information about the underlying current sources \cite{Dogandzic2000,Lukenhoner1998}. These models rely on a priori assumptions on the number of current sources and the propagation of secondary currents through the conductive tissues. In addition, they contain assumptions about the probability distribution of the background noise. One motivation behind our choice for a 3- component covariance model originates from the fact that the simpler model, that describes the covariance matrix as a Kronecker product of two components (spatial and temporal) has turned out to be useful for source localization. It has been shown, e.g. in  \cite{deMunck2002,Jun2006}, that current source estimation can be improved substantially by accounting for spatio-temporal noise covariances, which is done via the Kronecker product assumption. However, if stimulus induced experiments last for several minutes or more then the spatio-temporal model with two components may become questionable. 

%A physiological justification for such model is that it can be derived from the assumption that background activity is generated by randomly distributed dipolar sources for which amplitude time functions are independent of their locations \cite{deMunck1992}. The underlying cause for spatial correlations is that different sensors pick up signals from the same sources, and the temporal correlations reflect the fact that the current sources have an oscillatory nature. In evoked potential or evoked field experiments, where the same stimulus is presented hundreds of times, the random sources would reflect ongoing brain activity unrelated to the stimulus. Here the assumption can be made that the brain activity of interest follows the same spatio-temporal pattern after each stimulus. This means in particular that all trials have the same mean structure. The spatio-temporal covariance matrix describes the electro magnetic field distributions that are varying around the mean.

First, the assumption that the brain response is the same for each stimulus excludes the possibility of trial-to-trial variations or habituation effects \cite{David2006}. This aspect of response modeling has obtained some attention in the literature \cite{Limpiti2009,Sieluzycki2009}. A second limitation of the two domain model for the noise covariance is that it does not reflect the fact that the brain's state, such as subject's arousal, may change during the experiment. Such variations in brain states may modulate the amplitude variations that are superimposed on the brain activity induced by the stimulus. Since this aspect has been ignored until now, we will extend in the present paper the existing two domain covariance model by adding a third component which describes trial-to-trial variations in noise level.

Another aspect that was not taken into account in the two domain spatio-temporal covariance model is that the EEG/MEG signal is stationary on small time scales, which was, for instance, observed in \cite{Bijma2005}. We take care of this aspect by restricting the temporal component to be a Toeplitz matrix, \cite{Gray2001}. This restriction also reduces the number of unknown covariance parameters that need to be estimated.

Apart from being important for evoked potential and evoked field experiments, modeling of spontaneous EEG and MEG signals plays a role in resting state studies, where the subject is asked to stay awake while no particular stimulus is presented to the subject whose brain activity is recorded. Usually, such studies are performed using fMRI \cite{Lee2013,Guerra-Carrillo2014}, but because fMRI is an indirect measure of brain activity, combined EEG/fMRI experiments have become more popular, e.g.\ \cite{Goldman2002,Laufs2006}. 
Since EEG and fMRI are beneficial on complementary domains, integration of EEG and fMRI has become an active field of research.
Nevertheless, due to differences in dimensionalities of EEG and fMRI data, one of the largest challenges of a combined EEG/fMRI analysis is to develop mathematical models where both modalities are put in the same framework. Current practice is to assume a general linear model (GLM), in which the fMRI signal of a given voxel is the response variable and powers in different frequency bands extracted from EEG are used as regressors.  In this way, the use of the alpha activity derived in the occipital region has provided interesting novel insights in the generators of the alpha band \cite{Moosmann2003,Laufs2003} that would not have been possible with EEG alone. However, this approach suffers from the limitation that the choice of the regressor may only reflect a small part of the total EEG activity and the inclusion of other frequency bands, as explored in \cite{deMunck2009}, is only a partial solution. 

The proposed three domain covariance model provides a valuable alternative here because it relates the covariance to all electrodes instead of a small selected subset and no a priori choices about a certain frequency band need to be made. The physiological idea is that the EEG at rest is generated by randomly distributed and oscillating current sources of which the amplitude varies at the time scale at which the fMRI is recorded. The estimation of the parameters of the proposed model extracts the most dominant effects from the data. The estimated amplitude variations can be used for investigating EEG-fMRI correlations.

In Section 2 the mathematical details of the proposed model are presented and an estimation procedure is derived. A method for evaluating accuracy of the estimate is also presented. Section 3 contains simulation results, which assess the accuracy of the estimation procedure under the model and show which assumptions about the covariance components are related to increased accuracy. Moreover, real data results are presented for SEF MEG data and for EEG-fMRI data. In Section 4 the findings, extensions of the proposed framework and future applications are discussed.

\section{The model and estimation procedure}
\noindent
EEG/MEG data can be expressed as $X_{1},\ldots,X_{n}$, where $n$ is the sample size. We assume that each recording is written as $X_{k}=\left[X_{k}^{(1)},\ldots,X_{k}^{(r)}\right]$, where $X_{k}^{(d)}$ is the preprocessed $d$th epoch/trial organized in a 2-dimensional $p\times q$ matrix, such that $p$ represents the number of sensors and $q$ the number of time points within one epoch/trial. Although often $n=1$, the proposed method is applicable for general $n\geq 1$. 
Let vec be the operator that stacks columns of a matrix into one column in the order from first to last, so that vec$\left(X_{k}\right)$ is a vector of length $pqr$. We assume that vec$\left(X_{k}\right)$ has a multivariate normal distribution with mean $\mu$ and covariance matrix $\Sigma$ (vec$\left(X_{k}\right)\sim\mathcal{N}\left(\mu,\Sigma\right)$), and that $X_{1},\ldots,X_{n}$ are independent. Without loss of generality $\mu=\left(0,\ldots,0\right)$, so the signals oscillate around zero. Let $\Gamma$, $\Psi$ and $\Delta$ be positive definite matrices with dimensions $p\times p$, $q\times q$ and $r\times r$, respectively. Each of the matrices is related to one domain of the data: $\Gamma$ is the spatial component, $\Psi$ the temporal and $\Delta$ is the epochs/trials component. The relationship between the data and the three components is expressed by the assumption:  for $i_{1},i_{2}=1,\ldots,p$, $j_{1},j_{2}=1,\ldots,q$, $d_{1},d_{2}=1,\ldots,r$, Cov$\left(X_{k}^{(d_{1})}\left(i_{1},j_{1}\right),X_{k}^{(d_{2})}\left(i_{2},j_{2}\right)\right)=\Gamma\left(i_{1},i_{2}\right)\Psi\left(j_{1},j_{2}\right)\Delta\left(d_{1},d_{2}\right)$. The notation $A\left(i,j\right)$ stands for the $i,j$th element of the matrix $A$.
This assumption is equivalent to $\Sigma=\Delta\otimes\Psi\otimes\Gamma$, where $\otimes$ denotes the Kronecker product \cite{vanLoan}. For $\Gamma,\Psi,\Delta$ positive definite, $\Sigma=\Delta\otimes\Psi\otimes\Gamma$ is also positive definite. 

The next assumption is that the temporal covariances in $\Psi$ only depend on the time lag between time points, which is the same as assuming stationarity of the signal within one epoch/trial. Therefore the temporal matrix $\Psi$ is assumed to have a Toeplitz structure \cite{Xiao2012}, which means that elements that lie on the same subdiagonal are equal.
Moreover, we treat the data from different epochs as independent, which translates into the assumption that $\Delta$ is a diagonal matrix. 
In order to assure identifiability of $\Delta,\Psi,\Gamma$ from the Kronecker product, the constraints $\Gamma\left(1,1\right)=1$ and $\Delta\left(1,1\right)=1$ are added. 

In summary, the final model is
\begin{align}
\text{vec}\left(X_{1}\right),\ldots,\text{vec}\left(X_{n}\right)\sim\mathcal{N}\left(0,\Delta\otimes\Psi\otimes\Gamma\right),
\label{covariance_model}
\end{align} 
where vec$\left(X_{1}\right),\ldots,\text{vec}\left(X_{n}\right)$ are independent, $\Gamma$ is a $p\times p$ positive definite matrix, $\Psi$ is a $q\times q$ positive definite Toeplitz matrix, $\Delta$ is an $r\times r$ positive definite diagonal matrix, $\Gamma\left(1,1\right)=1$, and $\Delta\left(1,1\right)=1$.

Because the covariance matrix factorizes into a Kronecker product and two of the components have additional structure, the number of unknown parameters of $\Delta\otimes\Psi\otimes\Gamma$ equals $\frac{p\left(p+1\right)}{2}+q+r-2$, which is very low compared to the size of the data ($pqrn$). Thus one can expect to obtain accurate estimates of this structured covariance matrix if the model holds.

The covariance components $\Gamma,\Psi,\Delta$ are estimated by maximum likelihood. Even though there is no closed form estimator of the structured covariance matrix, likelihood equations for the matrices have been derived in \cite{vonRosen2012}. In such case each of the likelihood equations gives an expression of one Kronecker factor in terms of the others and, as a result, the likelihood equations cannot be solved analytically. Because we impose additional structure on $\Psi$ and $\Delta$, in our case the set of the likelihood equations becomes:

\begin{subequations}
\label{likelihood_equations}
\begin{align}
\Gamma &=\frac{1}{nqr}\sum_{k=1}^{n}{X_{k}
\left(\Delta^{-1}\otimes\Psi^{-1}\right)X_{k}^{T}},\label{eq_for_Gamma}\\
G&\left(\Psi^{-1}\sum_{k=1}^{n}{Y_{k}\left(\Delta^{-1}\otimes\Gamma^{-1}\right)Y_{k}^{T}\Psi^{-1}}-npr\Psi^{-1}\right)=\left(0,\ldots,0\right),\label{Toeplitz_equation}\\
\Delta& =\text{diag}\left(\frac{1}{npq}\sum_{k=1}^{n}{Z_{k}
\left(\Psi^{-1}\otimes\Gamma^{-1}\right)Z_{k}^{T}}\right),\label{eq_for_Delta}
\end{align}
\end{subequations}
with $\Gamma\left(1,1\right)=1$ and $\Delta\left(1,1\right)=1$. Here the matrices $Y_{k}$ and $Z_{k}$ contain the same elements as $X_{k}$, but arranged differently: $Y_{k}\left(j,p\left(d-1\right)+i\right)=Z_{k}\left(d,p\left(j-1\right)+i\right)=X_{k}^{(d)}\left(i,j\right)$, where $i=1,\ldots,p$, $j=1,\ldots,q$, $d=1,\ldots,r$. The dimensions of $Y_{k}$ and $Z_{k}$ are equal to $q\times pr$ and $r\times pq$, respectively. The function $G$  in \eqref{Toeplitz_equation} maps a $q\times q$ symmetric matrix into a $1\times q$ vector as follows: $G\left(A\right)=\left(G_{0}\left(A\right),\ldots,G_{q-1}\left(A\right)\right)$, where $G_{j}\left(A\right)=\sum_{i=1}^{q-1}{A\left(i,i+j\right)}$  is the sum of the elements of the matrix $A$ that lie on the $j$th subdiagonal. This function originates from the Toeplitz constraint that has been imposed on $\Psi$. Equation \eqref{Toeplitz_equation} can be derived from the likelihood equation for a Toeplitz covariance matrix by treating $\Gamma$ and $\Psi$ as if they were known, as is detailed in \ref{notation} and \ref{derivation_Toeplitz_equation}. Finally, the diag operator in equation \eqref{eq_for_Delta}  replaces the non-diagonal elements of a matrix by zeros. 

If no Toeplitz restrictions on $\Psi$ were imposed and $\Delta$ were unconstrained, a 3-component extension (\cite{vonRosen2012}) of the so-called iterative flip-flop algorithm (\cite{Dutilleul,Mardia}) could have been used to obtain numerical approximations of the ML-estimates of the covariance components. According to \cite{Gupta2011}, the necessary requirement for existence of the maximum likelihood estimator is $n\geq\max{\{\frac{p}{qr},\frac{q}{pr},\frac{r}{pq}\}}$. This condition guarantees that if the data is not rank deficient, each of the three likelihood equations provides a solution with respect to one covariance component that is positive definite conditional on the two other components being positive definite. Therefore if one uses the flip-flop algorithm and initial values of the two components $\Gamma,\Delta$ are positive definite, then all of the updates will stay positive definite. 

In our case the necessary condition on $n$ is different due to the Toeplitz assumption imposed on $\Psi$ and the diagonality of $\Delta$. The sample size required for estimating a Toeplitz covariance matrix is reduced by a factor two with respect to the sample size required for estimating an unrestricted covariance matrix \cite{Fuhrmann1988}. Therefore the requirement for a sample size that guarantees existence of a positive definite Toeplitz matrix $\Psi$ which satisfies \eqref{Toeplitz_equation} is also reduced by a factor two, which results in the new requirement: $n\geq\frac{\left \lceil{\frac{q}{2}}\right \rceil}{pr}$, where $\left \lceil{.}\right \rceil$ denotes the ceiling function. The diagonal matrix $\Delta$ at the left hand side of \eqref{eq_for_Delta} is guaranteed to be positive definite for any sample size and any configuration of the $p,q,r$ dimensions if the data have full rank, because diagonal elements of $\Delta$ are equal to sum of squares of scaled elements from $Z_{1},\ldots,Z_{n}$. Therefore under model \eqref{covariance_model} $n$ should satisfy
\begin{align}
\label{min_requirement}
n\geq\max{\{\frac{p}{qr},\frac{\left\lceil{\frac{q}{2}}\right\rceil}{pr}\}}.
\end{align}

The flip-flop algorithm needs to be adapted to take into account the additional assumptions about $\Psi$ and $\Delta$. The diagonality of $\Delta$ is easily taken care of by replacing the non-diagonal elements of each update of $\Delta$ by zeros, as follows from equation \eqref{eq_for_Delta}. 
Equation \eqref{Toeplitz_equation}, however, does not have a closed form solution for $\Psi$. Nevertheless \cite{Roberts2000} presents an iterative method for obtaining an approximation of the maximum likelihood estimate for a covariance matrix that has a Toeplitz structure. It is designed for the model in which there are independent realizations of $\mathcal{N}\left(0,T\right)$, where $T$ is a Toeplitz covariance matrix. The general idea presented in \cite{Roberts2000} to estimate $T$ by maximum likelihood is to embed it into a higher dimension circulant matrix. This is done by extracting the first $q$ columns and the first $q$ rows of the circulant matrix, where $q$ is the dimension of $T$.  Contrary to the case of Toeplitz matrices, for a circulant matrix $C$ a closed form maximum likelihood estimator exists that allows one to estimate $C$ directly from the data. A practical difficulty is, however, that the method of \cite{Roberts2000} requires an embedding dimension that is much higher than $q$. Therefore, large amounts of missing data have to be replaced by their expectations conditioned on the current estimate of $T$ and the original data. As a consequence, extensive simulations are necessary to determine the accuracy of this approach when a computationally acceptable embedding dimension is used. We adapt the procedure from \cite{Roberts2000} to make it possible to obtain approximate maximum likelihood estimates of $\Psi$ given the current updates of the estimates of $\Gamma$ and $\Delta$. A full description of the step of the algorithm in which $\Psi$ is updated is given in \ref{toeplitz_estimation}.

In our iterative estimation procedure the estimate of one component is updated based on the current estimates of the two other components, see Algorithm~\ref{alg:algorithm1}.

\begin{algorithm} 
Initialize: $\hat{\Gamma}_{0}=I_{p}$, $\hat{\Delta}_{0}=I_{r}$.

Repeat the following steps:
  \begin{algorithmic}[1]
    \State Calculate estimate $\hat{\Psi}_{m}$ of $\Psi$ based on $Y_{1},\ldots,Y_{n}$ and current estimates $\hat{\Gamma}_{m-1}$, $\hat{\Delta}_{m-1}$ of $\Gamma$ and $\Delta$ using embedding of $\Psi$ into a circulant matrix.
    \State Calculate estimate $\hat{\Delta}_{m}$ of $\Delta$ based on $Z_{1},\ldots,Z_{n}$, $\hat{\Gamma}_{m-1}$ and $\hat{\Psi}_{m}$ using \eqref{eq_for_Delta}.
    \State Calculate estimate $\hat{\Gamma}_{m}$ of $\Gamma$ based on $X_{1},\ldots,X_{n}$, $\hat{\Psi}_{m}$, and $\hat{\Delta}_{m}$ using \eqref{eq_for_Gamma}.
  \end{algorithmic} 
  \caption{}
  \label{alg:algorithm1}
\end{algorithm}

Due to the fact that updating the estimate of $\Psi$ is computationally complex, we also consider estimating $\Psi$ under a more general assumption, called persymmetry, which simplifies the estimation algorithm. A $q\times q$ matrix $A$ is a persymmetric matrix if it satisfies $JAJ=A^{T}$, where $J$ is the $q\times q$ matrix with ones on its anti-diagonal (the diagonal going from the lower left corner to the upper right corner), and zeros elsewhere. This means that a persymmetric matrix is symmetric with respect to its anti-diagonal. It is clear that a Toeplitz matrix is also persymmetric. Because there exists an analytical expression for the maximum likelihood estimator of a persymmetric covariance matrix \cite{Nitzberg}, we also consider this case. When the persymmetric assumption is made about $\Psi$, the first step in Algorithm~\ref{alg:algorithm1} is modified such that $\hat{\Psi}_{m}$ equals the maximum likelihood estimator for the persymmetric covariance matrix given $\hat{\Gamma}_{m-1}$ and $\hat{\Delta}_{m-1}$. It has the following form: $\hat{\Psi}_{m}=\frac{1}{2}\left(\hat{S}+J\hat{S}J\right)$, where $\hat{S}=\frac{1}{npr}\sum_{k=1}^{n}{Y_{k}\left(\hat{\Delta}_{m-1}^{-1}\otimes\hat{\Gamma}_{m-1}^{-1}\right)Y_{k}^{T}}$ and $J$ is defined above.

We perform simulation studies to assess the accuracy of the estimation procedure given above using data dimensions that are relevant for the present context. In these simulation studies not only the overall accuracy is measured, but also accuracies of the individual components of the covariance matrix. 

\subsection{Simulated data}
\label{simulations}
\noindent
The simulated data were generated from the $\mathcal{N}\left(0,\Delta\otimes\Psi\otimes\Gamma\right)$ distribution for $\Delta,\Psi,\Gamma$ positive definite, $\Psi$ Toeplitz matrix and $\Delta$ diagonal matrix. In the simulations physiologically relevant parameter values were taken as obtained from data analysis of MEG and EEG/fMRI recordings.

\subsection{Evaluation method}
\noindent
In simulation studies, the accuracy of our estimation method can be measured by making a direct comparison of the estimated covariance matrix $\hat{\Delta}\otimes\hat{\Psi}\otimes\hat{\Gamma}$ with the true covariance matrix $\Delta\otimes\Psi\otimes\Gamma$. We do this based on the relative mean squared error. In other words, the value of 
\begin{align}
\label{measure}
MSE = \frac{\|\hat{\Delta}\otimes\hat{\Psi}\otimes\hat{\Gamma}-\Delta\otimes\Psi\otimes\Gamma\|^{2}_{F}}{\|\Delta\otimes\Psi\otimes\Gamma\|^{2}_{F}}
\end{align} 
is calculated, where $\|\cdot\|_{F}$ denotes the Frobenius norm of a matrix. To check how well the individual covariance components are recovered, we use the relative mean squared errors for $\hat{\Gamma},\hat{\Psi},\hat{\Delta}$ separately:
\begin{subequations}
\label{MSE_components}
\begin{align}
MSE_{\hat{\Gamma}}&=\frac{\|\hat{\Gamma}-\Gamma\|_{F}^{2}}{\|\Gamma\|_{F}^{2}},\label{MSE_Gamma}\\
MSE_{\hat{\Psi}}&=\frac{\|\hat{\Psi}-\Psi\|_{F}^{2}}{\|\Psi\|_{F}^{2}},\label{MSE_Psi}\\
MSE_{\hat{\Delta}}&=\frac{\|\hat{\Delta}-\Delta\|_{F}^{2}}{\|\Delta\|_{F}^{2}}.\label{MSE_Delta}
\end{align}
\end{subequations}

We also assess the effect of taking into account the assumptions about $\Psi$ and $\Delta$ in the estimation procedure on the quality of the estimate. To do so, we estimate the components of the covariance matrix under modified assumptions about $\Psi$ or $\Delta$ and compare the accuracy of these estimates with the accuracy obtained under the correct assumptions about $\Psi$ and $\Delta$. Therefore each simulated data set, that is generated under model \eqref{covariance_model}, is used as input for estimation of $\Delta\otimes\Psi\otimes\Gamma$ under several sets of assumptions. 

We explain here which assumptions we use in the estimation and why. The assumption about the spatial component, $\Gamma$, is always that this matrix is positive definite, which is the weakest assumption that can be made. For the temporal component, we consider three alternative assumptions about $\Psi$ in estimation. The options are: positive definite Toeplitz, positive definite persymmetric and positive definite. The first possible assumption is the original assumption in model \eqref{covariance_model} and the third one is the most general restriction that can be made about the covariance component. The persymmetric assumption is a compromise between the two in terms of the number of unknown parameters and computational complexity of the estimation algorithm. The three alternative assumptions used in the estimation of the epoch/trials matrix component $\Delta$ are: positive definite diagonal, positive definite and identity assumption. While the diagonal assumption is the true assumption under \eqref{covariance_model}, $\Delta$ is also estimated under only positive definiteness to check how much the knowledge about its diagonal structure improves the accuracy of the estimator. $\Delta=I$ is considered to quantify the effect of incorporating trial-to-trial variations on the quality of the covariance estimator.

\subsection{Experimental data}
\subsubsection{MEG data}
\noindent
To illustrate the usefulness of our method, the model and estimation procedure were applied to an MEG data set recorded on a subject exposed to nervus medianus stimulation of the left wrist with pseudo random interstimulus intervals. The MEG was recorded on a whole head 151 channel CTF system, 3 bad channels were removed such that the data of $p=148$ gradiometers during the 30 minutes period at sampling frequency of 2084 Hz were considered in the analysis. 
The MEG recording was filtered using the order 1 Butterworth bandpass filter with frequency bands of 0.5Hz and 40Hz. No baseline correction was performed, as this would potentially compromise temporal stationarity \cite{Bijma2003}.
The original sampling frequency was decreased to 521 Hz by downsampling the signal. 
In the next step, the data that correspond to measurements recorded just after the occurrence of the stimulus were extracted. The window of 384 milliseconds, which corresponds to $q=200$ time points, was used. A common response function was assumed for all trials and the average response was subtracted from the MEG signal \cite{Bijma2005}. 

Because $p=148$ of the gradiometer signals  were taken and the stimulus occurred $r=509$ times during the recording, the dimension of the preprocessed data is $148\times 200\times 509$. Therefore the minimum requirement \eqref{min_requirement} for the sample size in the model with three positive definite covariance components is satisfied for $n=1$, which implies that analysis can be performed per subject.

\subsubsection{EEG data}
\noindent
We used the same EEG-fMRI data as were studied in \cite{deMunck2009}. 
Co-registered EEG-fMRI data were acquired from 16 healthy subjects (7 males, mean age 27, sd 9 years) while they lay rested in the scanner avoiding to fall asleep. Because the alpha activity is the most distinct aspect of the EEG signal for subjects that are awake and relaxed, we decided to select one subject with a large (L) and one with a small (S) amount of alpha oscillations. 

Due to the use of average reference, the rank of each observation written as a matrix with rows corresponding to channels, was decreased by 1. It means that the signal of one channel was fully determined by the signals of the remaining channels. Therefore a positive definite spatial component $\Gamma$ could not be estimated from these data. We therefore removed the central channel from the data after taking average reference.
%Because during the acquisition of fMRI, the magnetic field in the scanner changes each time a different slice %is acquired, the voltage fluctuations are induced that obscure the EEG signal. Even though the introduced %artifacts in EEG have an amplitude typically 100 times higher than the amplitude of the normal EEG signal, %the shape of the artifacts is approximately the same each time fMRI of the same slice is acquired. Therefore %this shape can be estimated by averaging and removed from the signal as was proposed in \cite{Allen2000}. 
%The fMRI artifact was removed by the method proposed in \cite{Allen2000}.
%Another source of artifacts in EEG is the blood movement under sensors which is related to cardiac cycles. %The amplitudes of these artifacts are lower than the artifacts caused by the fMRI scanner, but the shape %varies substantially for different heartbeats. This variability can be addressed by applying a hierarchical %clustering algorithm. The artifact template for each cluster is calculated by averaging and subtracted from %the EEG corresponding to the given cluster. Methods from \cite{deMunck2013}, that rely on the above %mentioned principles, were used for cleaning the EEG signal.

fMRI gradient and heartbeat artifacts and epochs with extreme outliers were removed. The EEG signal was filtered using a Butterworth bandpass filter with frequency bands of 0.5Hz and 40Hz. Because recordings from different epochs are assumed to be independent, the signal from the first 0.5 second and the last 0.5 second of each epoch was removed. The signal was downsampled such that 256 time points per epoch were obtained. Dimensions of the preprocessed data ($p\times q\times r$) were $62\times 256\times 569$ for subject L and $59\times 256\times 577$ for subject S.

\subsection{EEG-fMRI integration}
\noindent
The EEG regressor for the linear regression model is built from the estimate of the covariance matrix obtained from EEG under model \eqref{covariance_model}. Because diagonal elements of the estimate of $\Delta$ capture changes in the variance level of EEG during fMRI epochs and can be interpreted as measures of the amount of brain activity detected by EEG during particular fMRI epochs, we define the $\Delta$- regressor as the diagonal part of the estimate of $\Delta$. The values corresponding to isolated epochs that were removed were interpolated from the two values of the regressor from neighboring epochs. The regression model for fMRI with the $\Delta$- regressor is formulated in a similar way to \cite{deMunck2009}, wherein the alpha power variations of a selected set of electrodes were used and the hemodynamic response function was estimated for each voxel using time shifted copies of the alpha power regressor. Because the fMRI signal is also modulated by several non-neural phenomenons, like heartbeats and subjects's motion, regressors that are related to those aspects are added as confounders to the regression model. We first consider two regression models that only differ by the regressors of interest, which are either shifts of the $\Delta$- regressor or shifts of the alpha power regressor. Next, the regression model in which the regressors of interest are shifts of the $\Delta$- regressor with shifts of the alpha power regressor added to the set of confounders is used. For the three models significance of regressors of interest (shifts of $\Delta$- regressor or shifts of alpha power regressor) is tested by a partial F-test. To account for multiple testing, we control the false discovery rate (FDR) by applying the multiple testing correction of Benjamini and Hochberg \cite{Benjamini2005}. 

\subsection{Model validation}
\label{validation}
\noindent
The measure \eqref{measure} cannot be used for model validation for experimental data sets, because the true values of the parameters are unknown in such case. To assess how well model \eqref{covariance_model} fits the data, we estimated the covariance matrix under \eqref{covariance_model} using subselections of epochs/trials as follows. The set $\{1,\ldots,r\}$ of indices of epochs/trials was divided into 4 disjoint subsets of approximately equal sizes ($ind_{1}\cup ind_{2}\cup ind_{3}\cup ind_{4}=\{1,\ldots,r\}$) and data corresponding to each subset were used for estimation. The estimates were compared with the estimate obtained from the whole sample, $\hat{\Delta}\otimes\hat{\Psi}\otimes\hat{\Gamma}$, by: 
\begin{align}
\label{measure_evaluation}
\frac{\|\hat{\Delta}^{(i)}\otimes\hat{\Psi}^{(i)}\otimes\hat{\Gamma}^{(i)}-\hat{\Delta}_{ind_{i}}\otimes\hat{\Psi}\otimes\hat{\Gamma}\|_{F}^{2}}{\|\hat{\Delta}_{ind_{i}}\otimes\hat{\Psi}\otimes\hat{\Gamma}\|_{F}^{2}},
\end{align}
 where $\hat{\Delta}^{(i)}\otimes\hat{\Psi}^{(i)}\otimes\hat{\Gamma}^{(i)}$ denotes the estimate obtained from subsample $ind_{i}$ and $\hat{\Delta}_{ind_{i}}$ is equal to $\hat{\Delta}$ restricted to columns and rows with indices from $ind_{i}$.

The first check is whether different epochs/trials share the same spatio-temporal pattern. For this purpose, the subsets $ind_{1}$, $ind_{2}$, $ind_{3}$, $ind_{4}$ are generated randomly. This procedure is repeated 10 times, such that in total 40 values of \eqref{measure_evaluation} are be obtained. It is desirable that all of these values are low. 

The second check is whether there are no time-related changes of the spatio-temporal pattern not accounted for by $\Delta$. Now indices are subdivided such that approximately the first quarter of the sorted indices is the first subsample, the second quarter is the second subsample etc. 
%$\{1,\ldots,\lfloor{\frac{r}{4}}\rfloor\}$, $\{\lfloor{\frac{r}{4}}\rfloor+1,\ldots,2\lfloor{\frac{r}{4}}\rfloor\}$, $\{2\lfloor{\frac{r}{4}}\rfloor+1,\ldots,3\lfloor{\frac{r}{4}}\rfloor\}$, $\{3\lfloor{\frac{r}{4}}\rfloor+1,\ldots,r\}$.
Here, again, the measure \eqref{measure_evaluation} is used to assess the differences between estimates obtained from a subsample to the estimate acquired from the whole sample. It is desirable that all of these values are also low, because large values might indicate that the spatio-temporal covariance matrix $\Psi\otimes\Gamma$ is time-dependent.

\section{Results}
%We start this section by presenting simulation results, which assess performance of the estimation %procedure under our covariance model. The increased accuracy of each of the additional assumptions about %the components of the covariance matrix is also quantified. Additionally, accuracy improvement of the %introduced third covariance component in comparison with the model with the Kronecker product of two %components is measured. 
%Moreover, we present real data results with our covariance model applied to EEG and MEG data. The %components of the covariance matrix are estimated under model \eqref{covariance_model} and the model fit %is checked by comparing estimates obtained using subselections of epochs/trials. The estimator of the %covariance matrix of the EEG that was simultaneously collected with fMRI, is used in the EEG/fMRI regression %analysis. 
\subsection{Assessment of estimation procedure: simulation results}
\begin{figure}
\includegraphics[scale=0.32, trim=12mm 5mm 5mm 5mm]{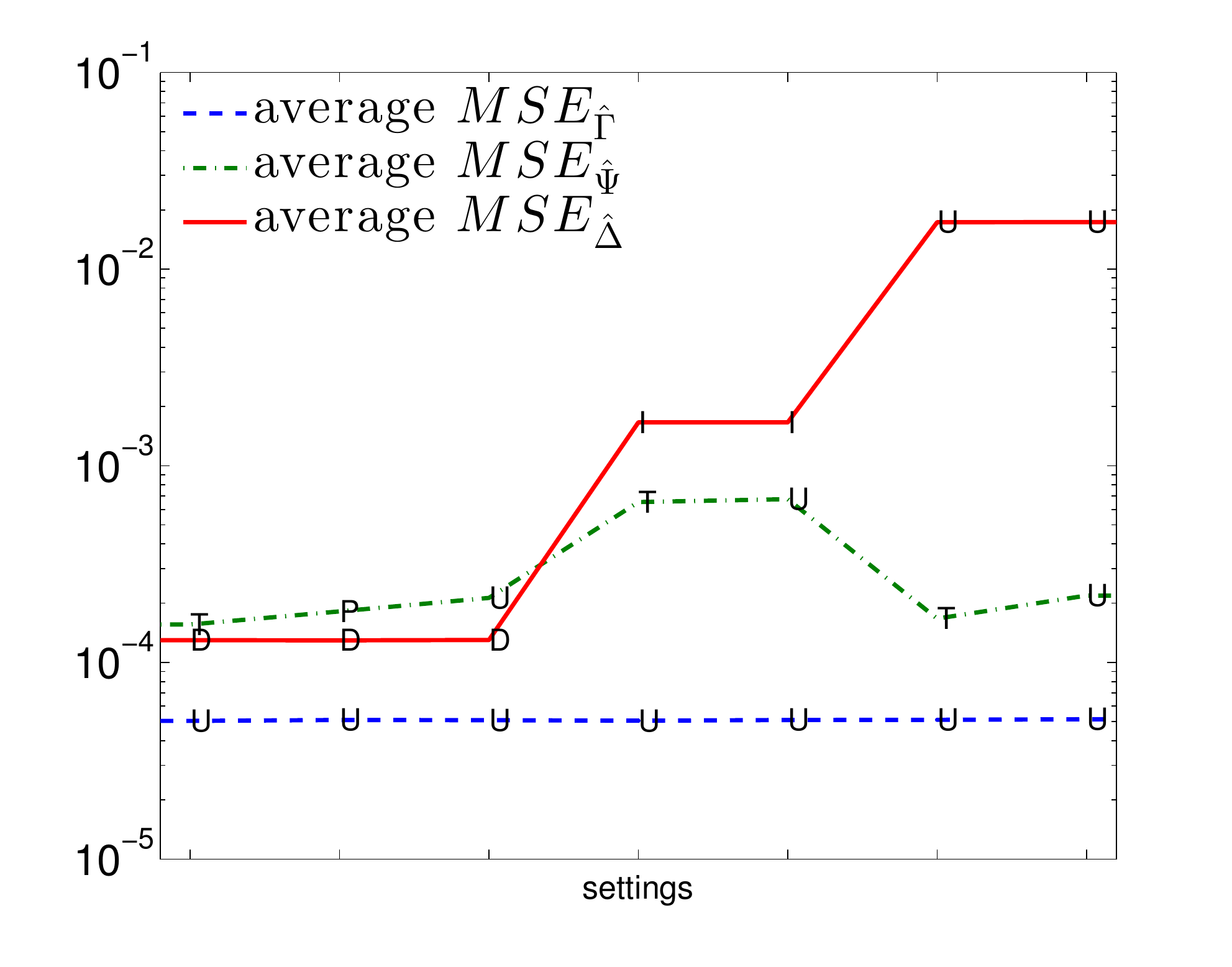}
\includegraphics[scale=0.32, trim=8mm 5mm 10mm 5mm]{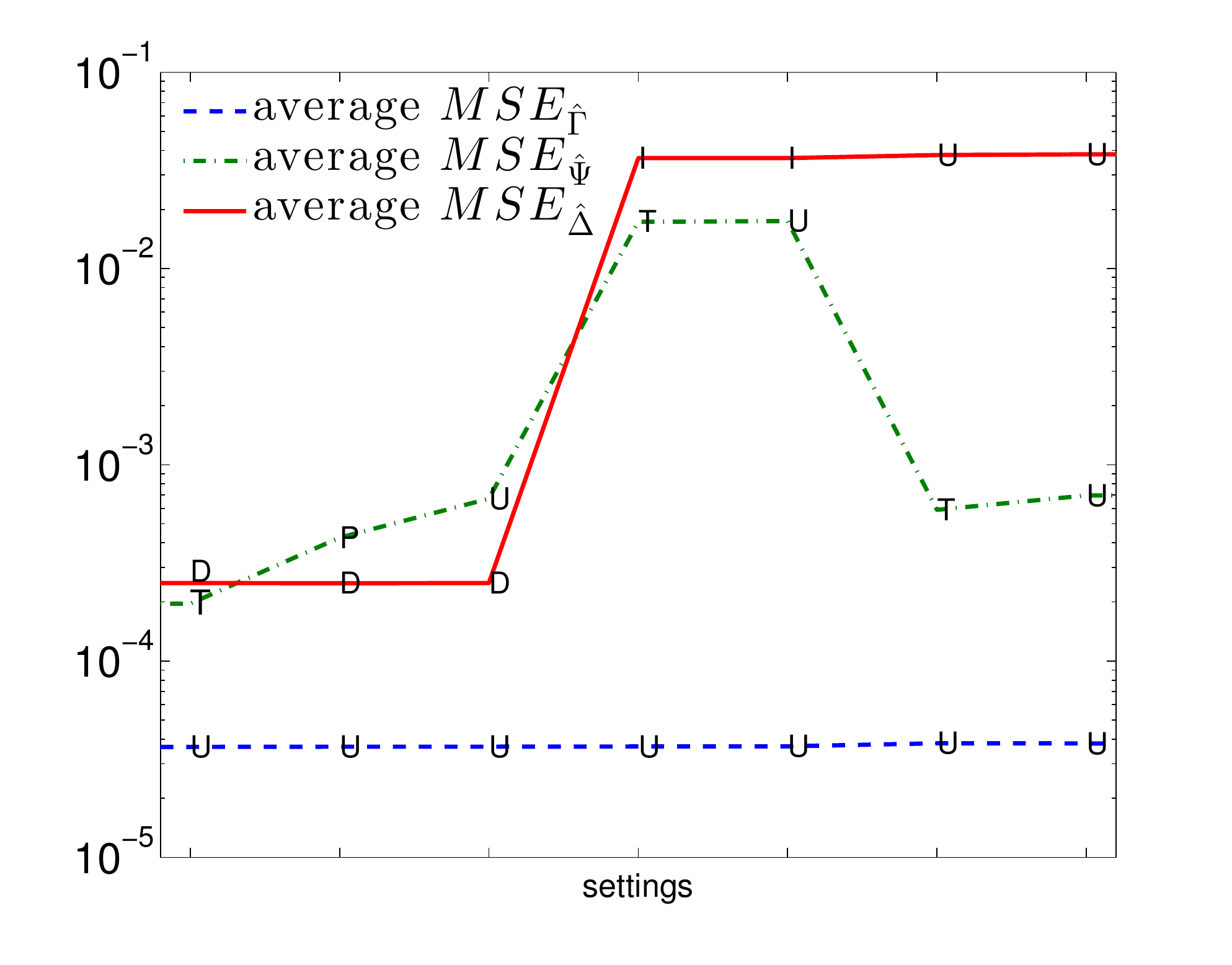}
\caption{Average (over 60 data sets) relative mean squared errors \eqref{MSE_components} of the estimates of components of the covariance matrix for simulated MEG (left) and EEG (right) data. Each value on a horizontal axis corresponds to one setting, which is coded by three letters that are added to plots.
Each letter that coincides with an MSE plot for a given component, denotes an assumption about the component that was used in estimation, where U is unrestricted, D is diagonal, T is Toeplitz, P is persymmetric and I is the identity matrix. The most left set of assumptions, $\left(U,T,D\right)$ for $\left(\Gamma,\Psi,\Delta\right)$ is the one of model \eqref{covariance_model}.}
\label{MSE_ingredients}
\end{figure}

\noindent
The "true" values for $\Gamma,\Psi,\Delta$, which satisfy the assumptions of model \eqref{covariance_model}, that were used in the simulations are the estimated values from the MEG and EEG experimental data analyses presented below (visualized in Figures \ref{matrices_MEG} and \ref{matrices_low_alpha}). For both studies the necessary requirement \eqref{min_requirement} regarding the sample size in relation to the dimensions of the covariance components was satisfied. A number of 60 data sets were simulated for EEG and MEG. Each sample, which was simulated under model \eqref{covariance_model}, was used for estimation under 7 different sets of assumptions about the components. For each estimated covariance matrix the relative mean squared error \eqref{measure} and relative mean squared errors of individual components of the covariance matrix \eqref{MSE_components} were calculated. Means of $MSE$s for each set of assumptions are given in Table \ref{assumptions_influence} and means of $MSE_{\hat{\Gamma}}$s, $MSE_{\hat{\Psi}}$s, and $MSE_{\hat{\Delta}}$s are given in Figure \ref{MSE_ingredients}. The covariance matrix and its components were most accurately recovered if all assumptions from \eqref{covariance_model} were taken into account in the estimation.
\begin{center}
\begin{table}
\caption{Average relative mean squared errors \eqref{measure} of the covariance estimates obtained from samples that were simulated under \eqref{covariance_model}, together with percentages of the mean squared error obtained under \eqref{covariance_model}. Each sample was used in covariance estimation of $\Delta\otimes\Psi\otimes\Gamma$ under several sets of assumptions about $\Gamma,\Psi,\Delta$. Assumptions that were used are coded by letters, where U is unrestricted, D is diagonal, T is Toeplitz, P is persymmetric and I is the identity matrix.}
\label{assumptions_influence}
\begin{tabular}{|l|l|C|C|C|l|} 
\hline
input&dimensions&\multicolumn{3}{c|}{\multirow{1}{*}{assumptions about}}&average MSE\\
type&of each&\multicolumn{3}{c|}{\multirow{1}{*}{the components used}}&according to \eqref{measure}\\
&data set&\multicolumn{3}{c|}{\multirow{1}{*}{in the estimation}}&\\
&&$\Gamma$:&$\Psi$:&$\Delta$:&\\
\hline
MEG&$p=148$&U&T&D&1.3e-04 (100\%)\\
\cline{3-6}
&$q=200$&U&P&D&1.5e-04 (115\%)\\
\cline{3-6}
&$r=509$&U&U&D&1.8e-04 (139\%)\\
\cline{3-6}
&$n=1$&U&T&I&1.1e-03 (846\%)\\
\cline{3-6}
&&U&U&I&1.2e-03 (924\%)\\
\cline{3-6}
&&U&T&U&1.72e-02 (1324\%)\\
\cline{3-6}
&&U&U&U&1.74e-02 (1339\%)\\
\hline
EEG&$p=59$&U&T&D&2e-04 (100\%)\\
\cline{3-6}
&$q=256$&U&P&D&4e-04 (200\%)\\
\cline{3-6}
&$r=577$&U&U&D&6.8e-04 (340\%)\\
\cline{3-6}
&$n=1$&U&T&I&1.4e-02 (7000\%)\\
\cline{3-6}
&&U&U&I&1.5e-02 (7500\%)\\
\cline{3-6}
&&U&T&U&3.7e-02 (18500\%)\\
\cline{3-6}
&&U&U&U&3.9e-02 (19500\%)\\
\hline
\end{tabular}
\end{table}
\end{center}
%Stability of the linear regression model with our regressor is checked by fitting the model on the subsets of %the data. 
%The dimensions used for the simulation studies are taken from the preprocessed task induced MEG data and %spontaneous EEG data which was simultaneously recorded with fMRI. 
\subsection{Experimental data results}
\subsection*{Regarding MEG}
\begin{figure}
\begin{center}
\includegraphics[scale=0.45]{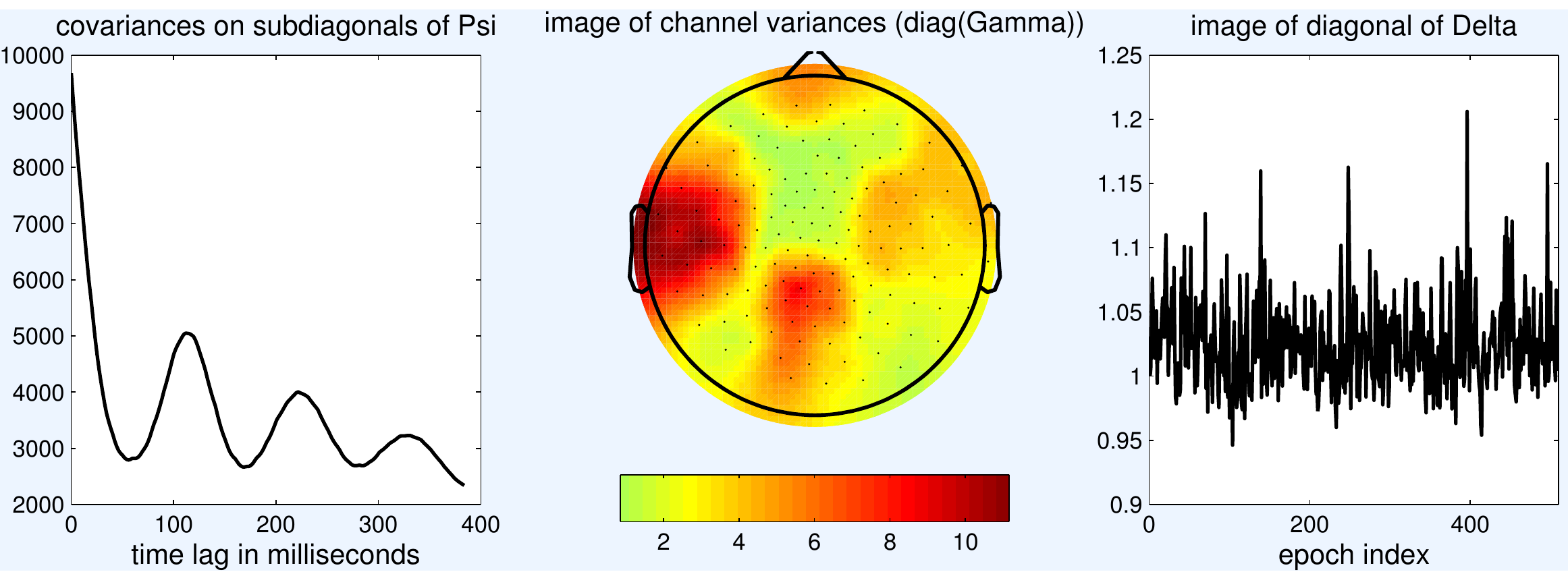}
\includegraphics[scale=0.38]{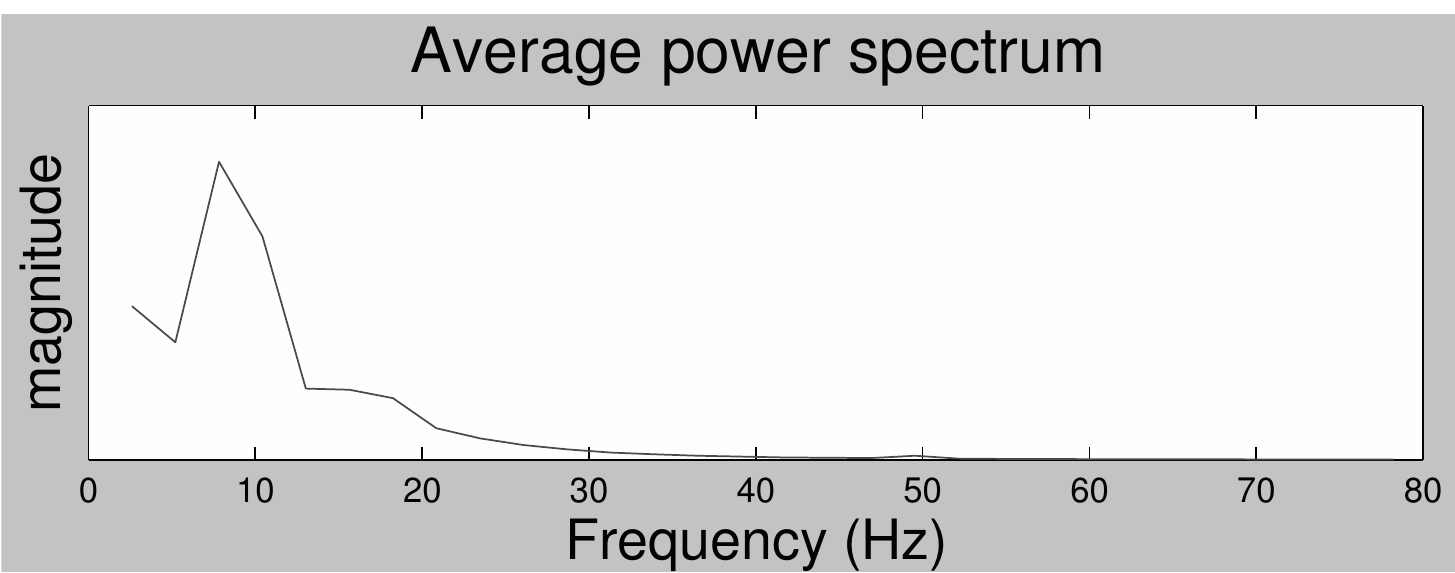}
\caption{Components of the covariance matrix estimated from the noise of the SEF data. Upper-left: covariances on subdiagonals of the temporal component. Upper-middle: variances of the MEG channels (diagonal elements of the estimate $\Gamma$) projected onto the head model. Values in between channels were interpolated. Upper-right: estimates of trial-to-trial variations (diagonal entries of the estimate of $\Delta$). Lower: estimate of the average power spectrum.}
\label{matrices_MEG}
\end{center}
\end{figure}

% filter out frequencies <0.5Hz and >40Hz
% remove boundary effects
% extract post-stimulus intervals
% estimate response and subtract it from the data from each interval
\noindent
 Model \eqref{covariance_model} was assumed for the preprocessed SEF MEG trials of one subject. The estimates of the covariance components $\hat{\Gamma},\hat{\Psi},\hat{\Delta}$ were determined using our iterative procedure. The results are visualized in Figure \ref{matrices_MEG}. Temporal covariances show decaying oscillations of approximately 10Hz. This is the effect of alpha activity, which was dominant in the MEG of this subject, as indicated by the highest peak in the estimate of the average power spectrum. Higher spatial variances are on the left side of the brain. This can be related to the fact that the right side of the brain was stimulated. Estimates of trial-to-trial variations, which are also visualized, show little variability for this subject.

The model validation was done according to Section \ref{validation}. For the random subselection of trials, the average of \eqref{measure_evaluation} was $0.0136$ and the standard deviation $0.015$. If subsets of data were created from consecutive trials, the values of \eqref{measure_evaluation} were equal to $0.020, 0.003, 0.013, 0.005$. These values are very low, which means that the spatio-temporal covariance $\Psi\otimes\Gamma$ does not vary much over time.

\subsection*{Regarding EEG}
\begin{figure}
\begin{center}
\includegraphics[scale=0.45]{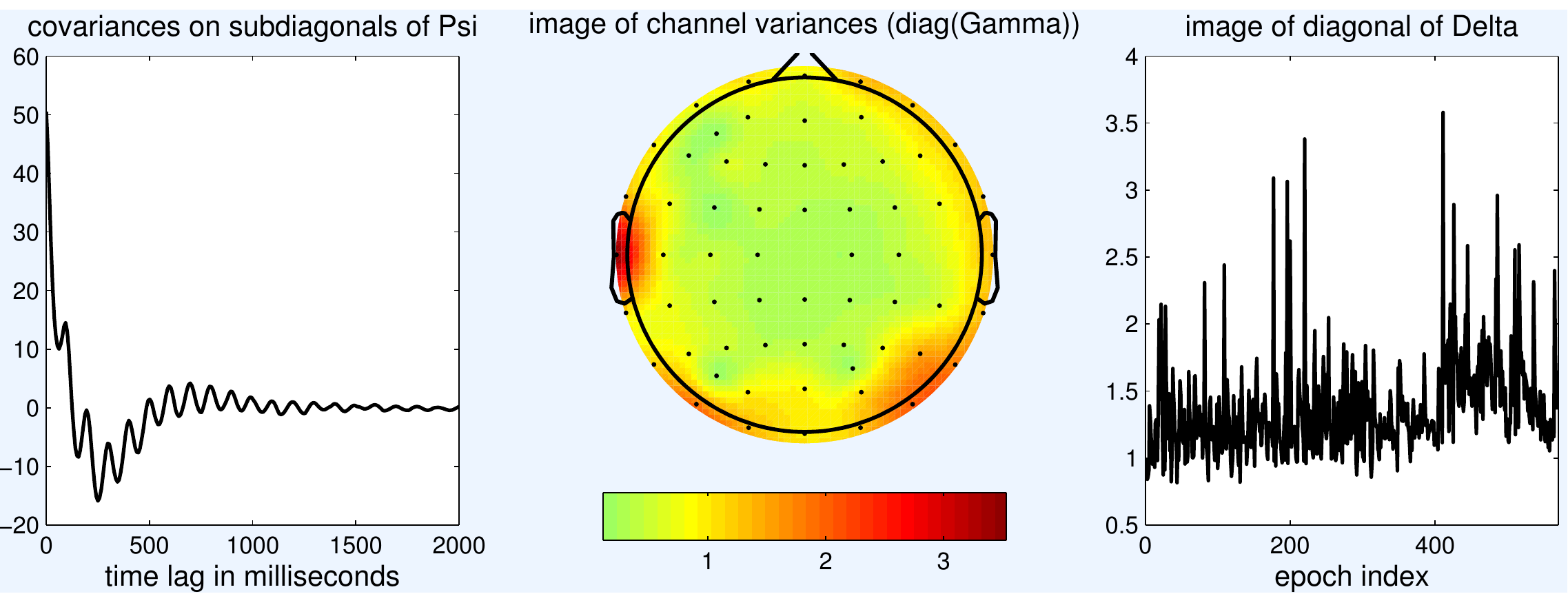}
\includegraphics[scale=0.38]{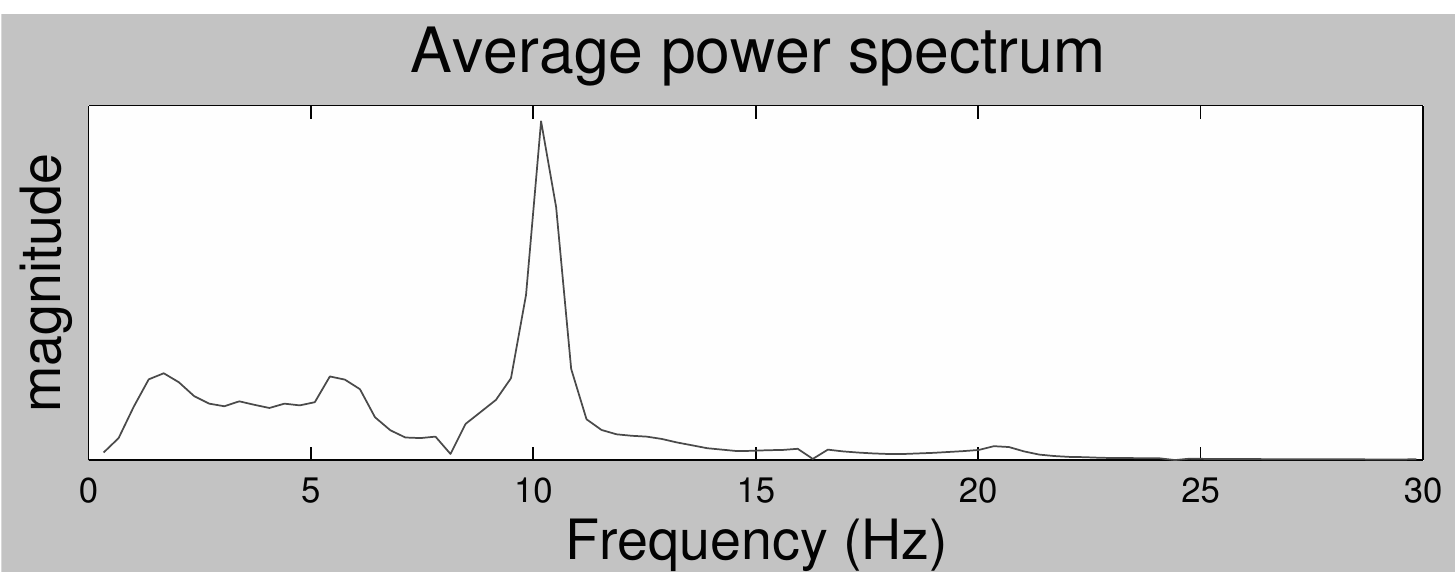}
\caption{Components of the covariance matrix estimated from EEG of subject L, visualized in the same way as in Figure \ref{matrices_MEG} for MEG data.}
\label{matrices_EEG_high_alpha}
\end{center}
\end{figure}

\begin{figure}
\begin{center}
\includegraphics[scale=0.45]{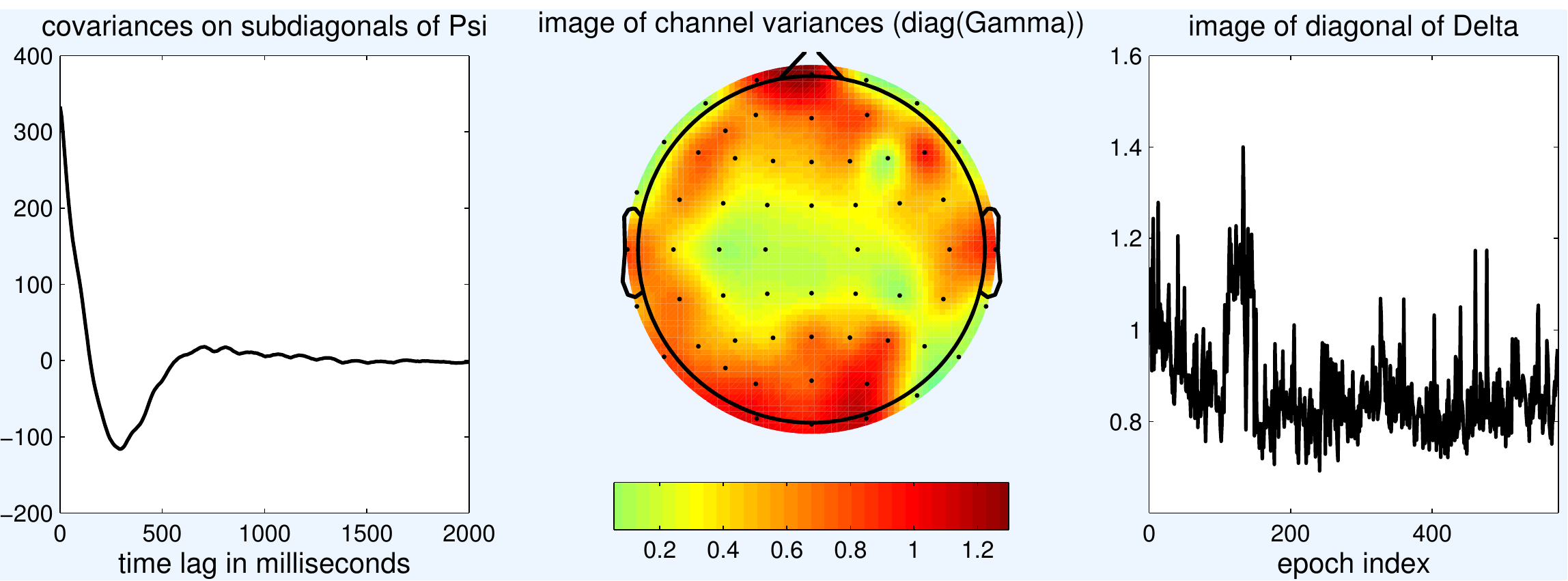}
\includegraphics[scale=0.38]{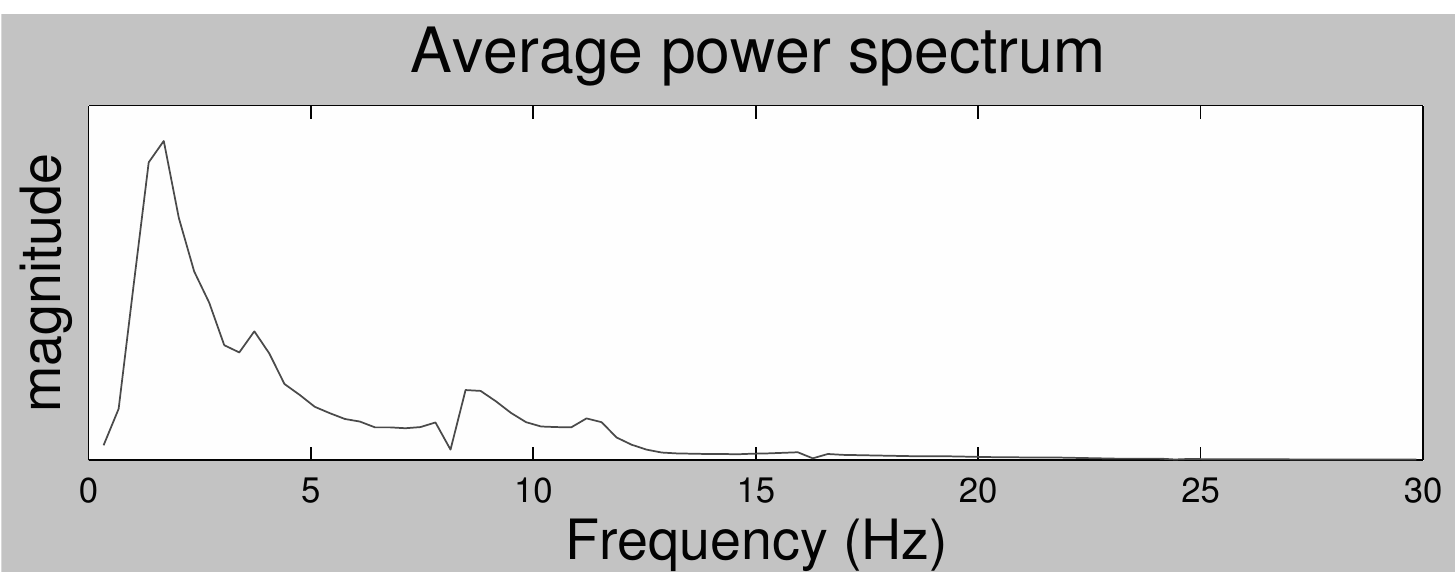}
\caption{Components of the covariance matrix estimated from EEG of subject S. For comparison, the estimated components for a subject with a large amount of alpha activity are given in Figure \ref{matrices_EEG_high_alpha}.}
\label{matrices_low_alpha}
\end{center}
\end{figure}

\noindent
Preprocessed EEG data of subjects S and L were used as inputs for individual covariance estimation under \eqref{covariance_model}. Results for subject L are visualized in Figure \ref{matrices_EEG_high_alpha} and for subject S in Figure \ref{matrices_low_alpha}. These figures show the estimates of spatial, temporal and epoch-related covariance components. The amount of alpha activity, that is expressed by the shape of the estimated power spectrum, is reflected by the estimate of the temporal covariance component: in Figure \ref{matrices_EEG_high_alpha} there are clear alpha oscillations, while in Figure \ref{matrices_low_alpha} such oscillations are absent.

The model validation was done for both subjects, again according to Section \ref{validation}. For subject S for the random subselection of epochs the average of \eqref{measure_evaluation} was $0.0016$ and the standard deviation $0.0012$, and for the consecutive subselection of epochs the values of \eqref{measure_evaluation} were equal to $0.054, 0.020, 0.011, 0.044$. These values are low, which suggests that the spatio-temporal covariance pattern was captured by the model. The results for subject L were similar.

\subsection*{Regarding integration of EEG-fMRI}
\begin{figure}
\begin{center}

\includegraphics[height=25mm]{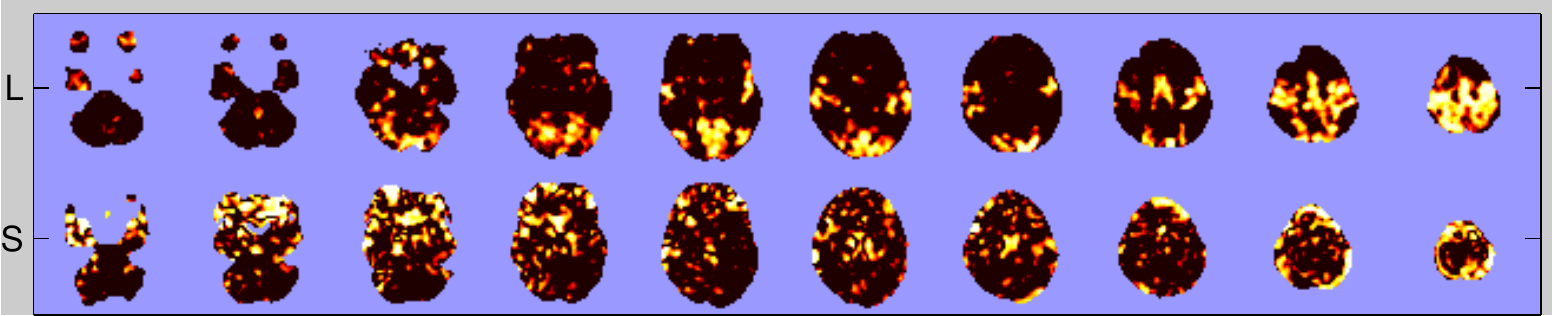}
\vspace{0.3cm}

\includegraphics[height=25mm]{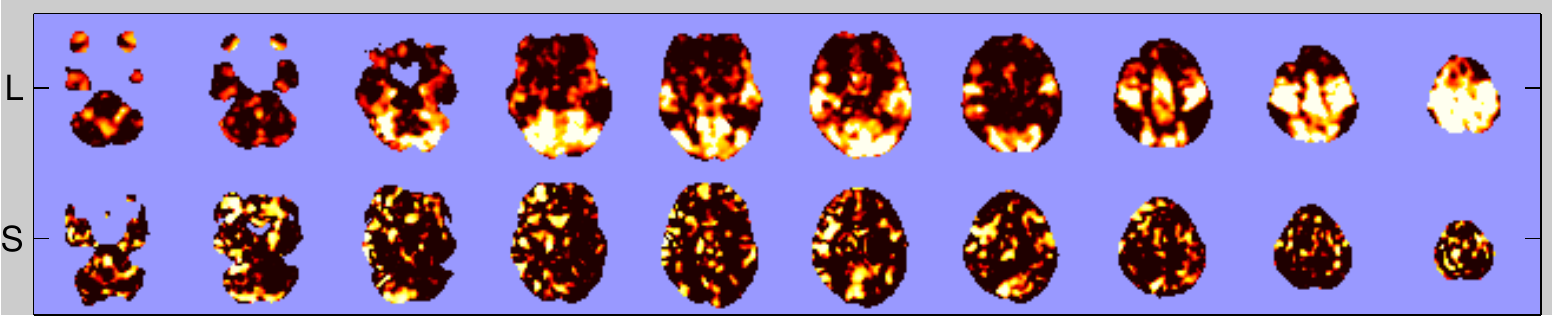}
\caption{Upper, from red to yellow: voxels for which shifts of the $\Delta$- regressor are significant at FDR=0.01. Lower, from red to yellow: voxels for which shifts of the alpha power regressor are significant at FDR=0.01. Subjects are marked on Y axes, where L denotes the subject with large, and S with small amount of alpha activity. The lighter colors indicate more significant voxels than the darker colors within one subject and one group of regressors.}
\label{KP_alpha_sig_FDR_0.01}
\end{center}
\end{figure}

\noindent
Significance of the two sets of regressors of interest (shifts of the $\Delta$- regressor and shifts of the alpha power regressor) was tested by the partial F-test conditionally on the same set of confounders with FDR=0.01. Since the test was done for each voxel, those voxels for which regressors of interest were significant, are indicated in the axial brain slices in Figure \ref{KP_alpha_sig_FDR_0.01}. We use the term significant voxels, which denotes voxels for which regressors of interest turned out to be significant for FDR=0.01.
There are noticeable differences in the number of significant voxels between subjects and also between the two models. For subject L there are more significant voxels when the alpha power regressor is used. Nevertheless the areas occupied by such voxels for the two models are quite similar. For subject S the situation is different. Namely, the number of significant voxels is similar, but their location is different for the two models. If the alpha power regressor is used, significant voxels are mostly located in the occipital lobe while if the $\Delta$- regressor is used, many significant voxels are in the frontal lobe. 

%Due to large variability in the number of detected voxels for different subjects and models, 3000 voxels with the lowest p values for each of the two subjects are plotted in Figure %\ref{KP_alpha_sig_min_3000}. These plots confirm our earlier observations about the areas in the brain that are detected for two models and two subjects.
%\begin{figure}
%\begin{center}
%\includegraphics[height=18mm]{plots/KP_sign_5_10_no_alpha_min_3000_.pdf}
%\includegraphics[height=18mm]{plots/alpha_sign_5_10_no_KP_min_3000_.pdf}
%\caption{Upper, in white: 3000 voxels for which the partial F test for our regressors resulted in the lowest p values, for subjects 5 and 10, Lower, in white: 3000 voxels with the lowest p %values in the partial F test applied for the alpha regressors, for subjects 5 and 10.}
%\label{KP_alpha_sig_min_3000}
%\end{center}
%\end{figure}

\begin{figure}
\begin{center}
\includegraphics[height=25mm]{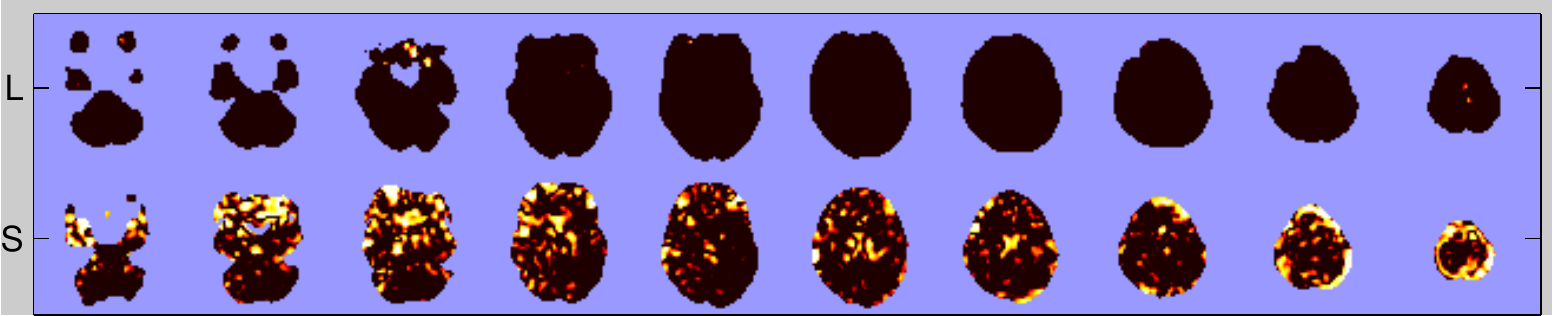}
\caption{From red to yellow: voxels for which the partial F test for shifts of $\Delta$- regressor conditionally on confounders and shifts of alpha regressor was rejected at FDR=0.01. Upper figures represent the results for subject L and lower for subject S. The brain areas from red to yellow correspond to significant voxels. The lighter colors indicate more significant voxels than the darker colors within one subject.}
\label{KP_cond_alpha}
\end{center}
\end{figure}

In order to localize the regions for which addition of shifts of the $\Delta$- regressor to the model with shifts of the alpha power regressor (and confounders) significantly improves the model,
the partial F-test at FDR=0.01 was applied for the shifts of the $\Delta$- regressor, conditionally on the confounders and shifts of the alpha power regressor, see Figure \ref{KP_cond_alpha}.
The results show that there are hardly any significant voxels for subject L. For subject  S, however, many voxels are significant, which are mostly located in the frontal regions. 

\begin{figure}
\begin{center}
\includegraphics[height=40mm]{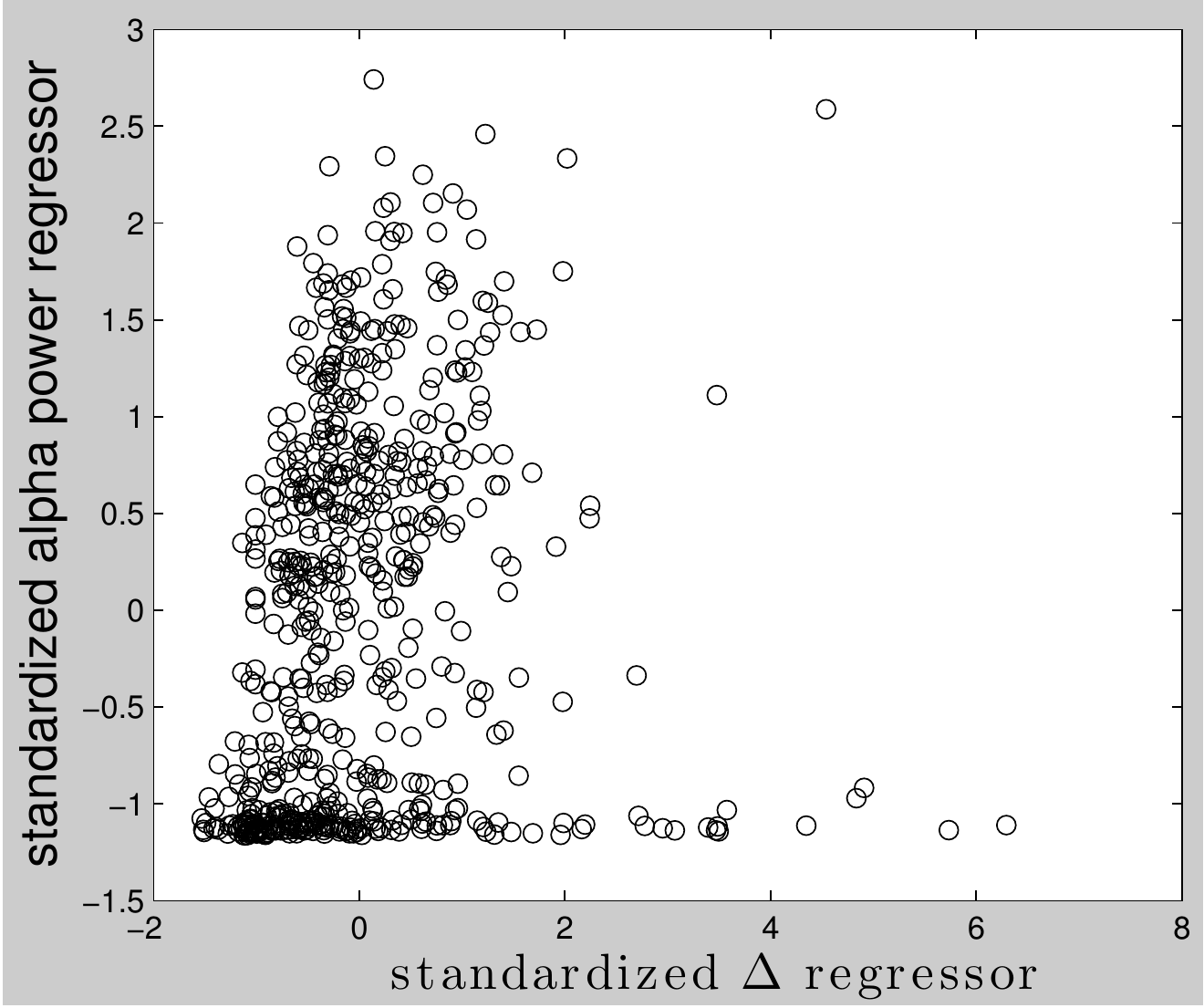}
\includegraphics[height=40mm]{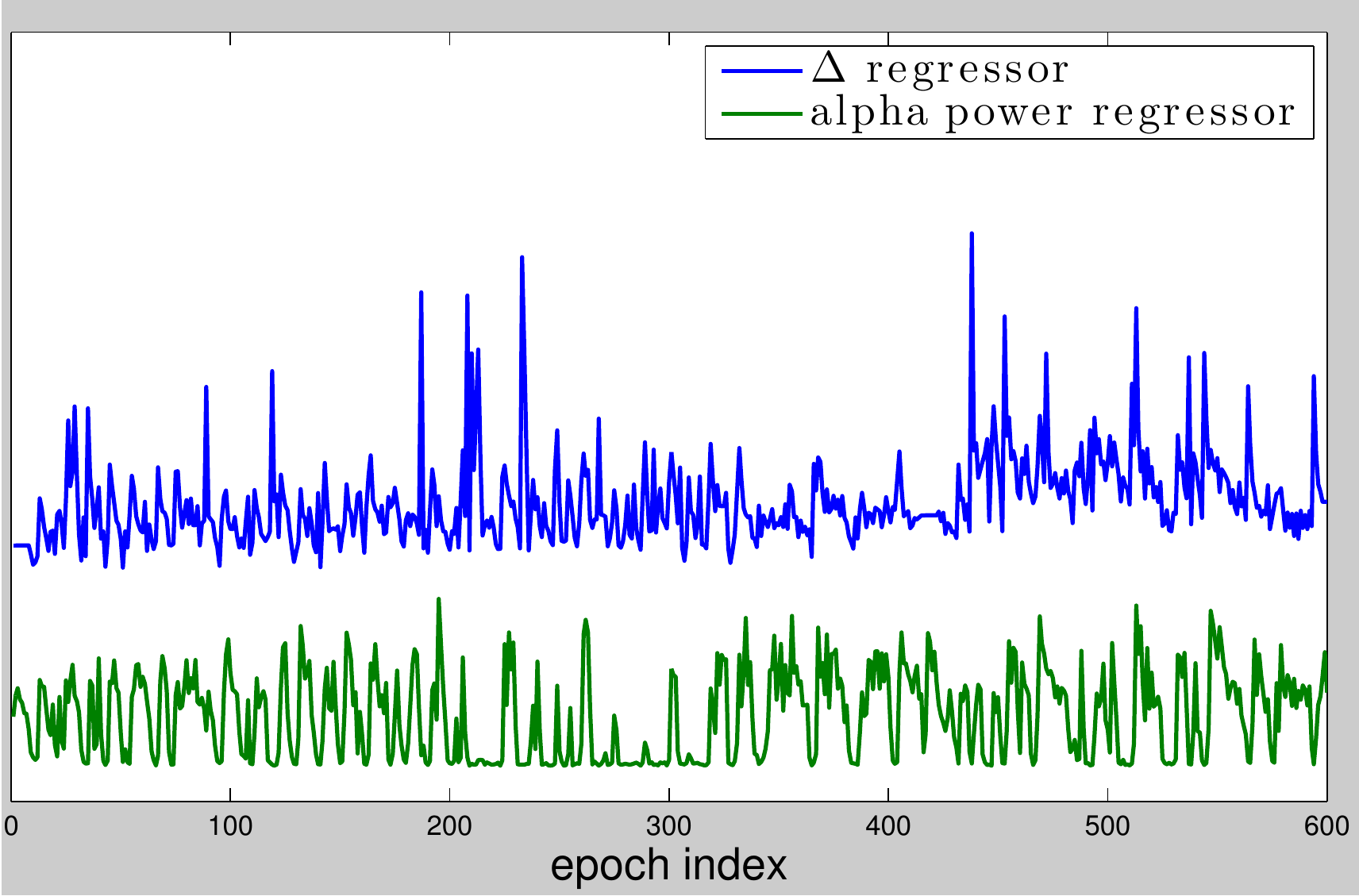}
\caption{Comparison of $\Delta$- regressor and alpha regressor extracted from EEG of subject L. Left: scatter plot of the two regressors. Right: time series plots of the two regressors; upper: $\Delta$ regressor, lower: alpha power regressor.}
\label{regressors_comparison_high_alpha}
\end{center}
\end{figure}

\begin{figure}
\begin{center}
\includegraphics[height=40mm]{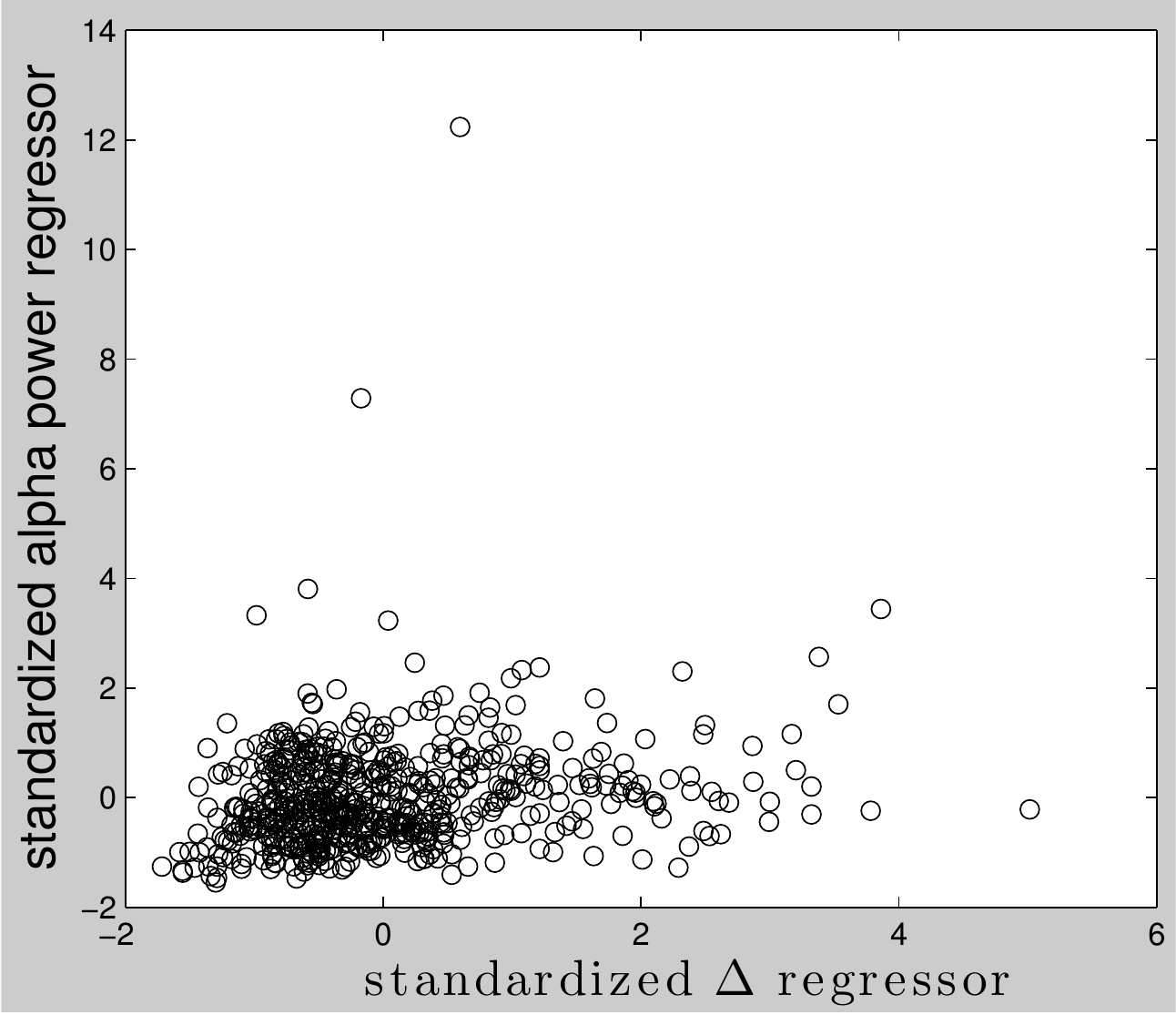}
\includegraphics[height=40mm]{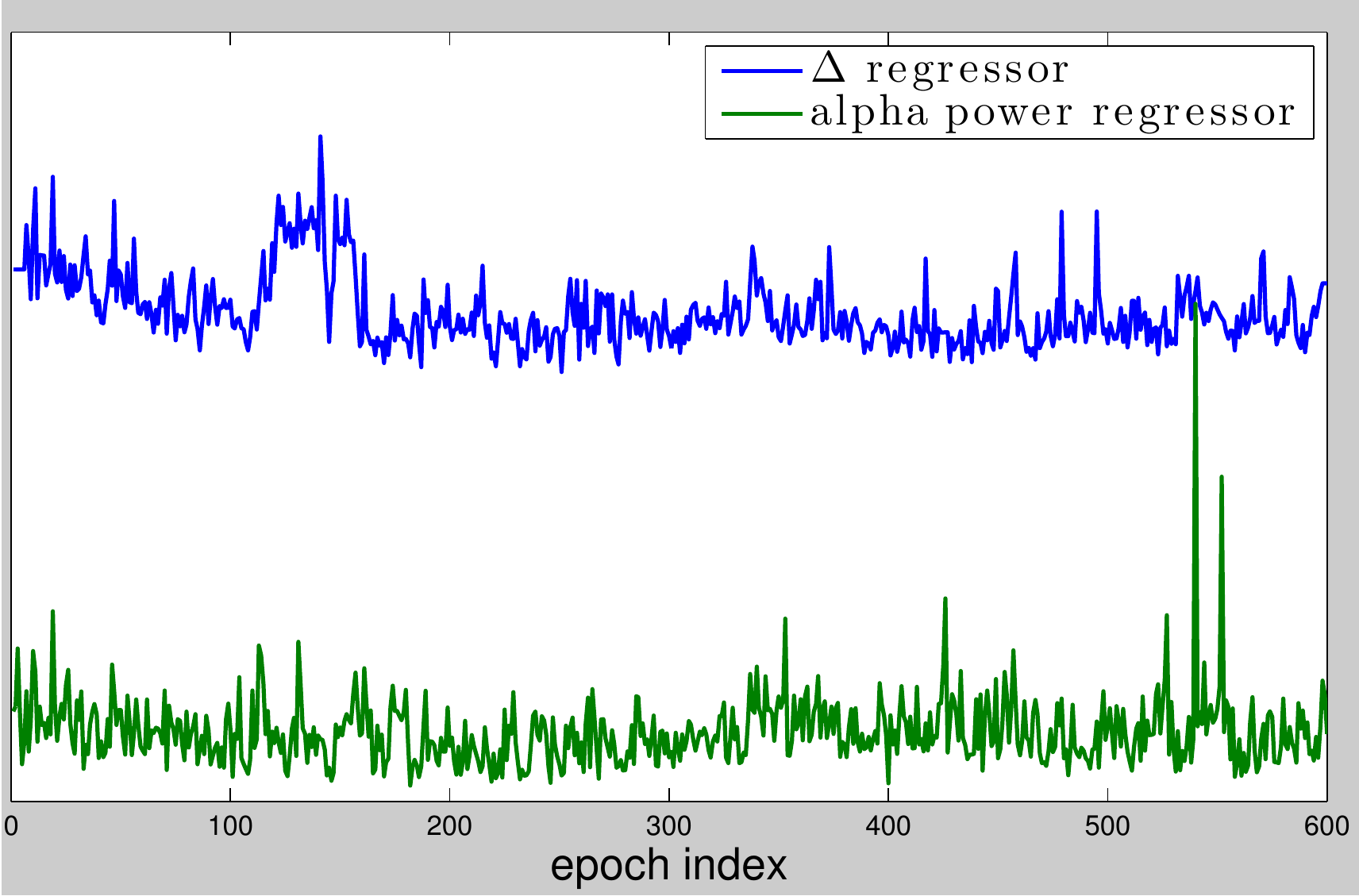}
\caption{The same comparison of the two regressors as in Figure \ref{regressors_comparison_high_alpha} , but here for a subject S.}
\label{regressors_comparison_low_alpha}
\end{center}
\end{figure}

We made a direct comparison of shapes of the alpha regressor and the $\Delta$- regressor for both subjects, see Figures \ref{regressors_comparison_high_alpha} and \ref{regressors_comparison_low_alpha}. It appears that there is a more pronounced difference between the two sets of regressors for subject S, as Spearman's rank correlation coefficient between the two regressors equals 0.218 for this subject and 0.316 for subject L.


\section{Discussion and conclusions}
\noindent
The introduced model for the covariance of EEG/MEG incorporates local stationarity of the data and variations in the noise level over epochs/trials. These aspects are visible in the data, but are not taken into account in the earlier approaches that used as covariance model a Kronecker product of two components e.g.\ \cite{deMunck2002}. Therefore in this study the local stationarity was explicitly imposed by assuming that the temporal covariance component has a Toeplitz structure. Moreover, a third component was added into the Kronecker product, describing the amplitude variations of the brain's background activity. These model extensions contribute to a more precise description of the underlying physiology of EEG and MEG signal generation. For instance in \cite{Linkenkaer2004} evidence has been found that the amplitude of the background activity predicts the chance that a stimulus is consciously perceived by the subject. Also other studies suggest that the brain's background activity is not simply additive noise \cite{Huizenga1995}. However, in order to test such hypothesis in a formal statistical setting, accurate mathematical models are needed and we consider our three domain covariance model as a next step towards such models.  

The parameterization of the covariance matrix via the Kronecker product of three matrices and the assumptions imposed on these matrices provide a great dimension reduction of the unknown parameters and make it possible to estimate the covariance matrix per subject ($n=1$). It needs to be stated that only partial theoretical results are available in the literature about the properties (existence, uniqueness and accuracy) of the maximum likelihood estimator of the covariance matrix that is a Kronecker product of two or more components \cite{RosEtAl}. We performed simulation studies based on the data that resemble EEG/MEG, with $n=1$, to assess the accuracy of the estimation procedure if model \eqref{covariance_model} holds. Results clearly show that the estimates are very close to the true covariance matrix that was used for data generation. Therefore one can expect to obtain reliable estimation results if our model fits the data. 

Another aspect that was investigated using simulated data from model \eqref{covariance_model} is the influence of incorporating variations in the noise level over trials and stationarity of the signal within each trial in the estimation, given that there is variation in noise level and stationarity of the signal. It was assessed how accuracy of the covariance estimator changes if each of these assumptions is dropped in the estimation.
Moreover, the estimate under model \eqref{covariance_model} was compared to the estimate if estimation was done assuming no trial-to-trial variations. The results show that correct assumptions about the components of the covariance matrix in the estimation procedure provide the most accurate estimate. It turns out that the Toeplitz assumption about $\Psi$ does not have a strong effect on the accuracy of the covariance estimate. However, a high drop in accuracy is observed when no trial-to-trial variations are assumed, i.e. when $\Delta=I$ or when diagonality of $\Delta$ is skipped. The gain in accuracy related to estimating the trial-to-trial variations is higher for simulated EEG data than for simulated MEG data. This can be explained by the fact that higher trial-to-trial variations are observed in our EEG sample than in our MEG sample. 

Accuracies of estimates of the individual components of the covariance matrix were determined for all simulated samples and under each set of assumptions. It turns out that the quality of $\hat{\Gamma}$ is comparable under each set of assumptions. However, the accuracy of $\hat{\Psi}$ can be affected by the assumption about $\Delta$, which happens when no trial-to-trial variations are assumed. In such case the accuracy of both $\hat{\Psi}$ and $\hat{\Delta}$ decreases.

One could argue that the amount of variation per epoch could be estimated by simply treating signals from all channels and the same epoch as realizations of the univariate normal distribution with unknown variance parameter and estimating this parameter. Nevertheless, such an approach is not recommended, because the estimates would also capture variability of the most varying channels. In our approach $\Delta$ represents the activity in different epochs corrected for temporal and spatial effects that are intertwined with these measures.

The method of estimating $\Psi$ under the Toeplitz assumption requires a certain size of the circulant extension to assure convergence to the maximum likelihood estimator. Satisfying this requirement, however, is not feasible in practice on a single machine and it would lead to increased memory consumption and computation time on a multicore machine. For these reasons we take one of the smallest possible sizes of the extension, $l=2q-1$. We infer from our simulation studies that this size of the circulant extension provides an accurate approximation of the covariance estimate. Nevertheless, there exists a way of simplifying the estimation step of $\Psi$ in terms of computational complexity and computational time. This is achieved by relaxing the Toeplitz assumption about $\Psi$ and imposing a weaker, persymmetric, assumption. In such case the likelihood equation for $\Psi$, which would replace \eqref{Toeplitz_equation}, can be solved analytically. However, one needs to be aware that estimating $\Psi$ under the persymmetric assumption is associated with a decreased accuracy of the estimate of $\Psi$ if the true $\Psi$ is a Toeplitz matrix.

Diagonality of the newly introduced covariance component reflects independence between different epochs/trials. This a natural assumption for the stimulus evoked data, because the post-stimulus trials are extracted from the continuous signal such that the time points that correspond to different trials are separated by hundreds of milliseconds. On the other hand, from EEG/fMRI experiment, the EEG signal is naturally divided into epochs that are defined by the fMRI scans. Hence, part of the data in between each two consecutive epochs needs to be removed to assure independence between epochs. Fortunately, since dependence over time is short, the removed part can be small.

%It turns out that making the Toeplitz assumption about $\Psi$ and the diagonal assumption about $\Delta$ %is neither arbitrary, nor unreasonable. We used spontaneous EEG and task induced MEG recording of one %subject, therefore $n=1$, as input to the estimation procedure. If model \ref{covariance_model} was %assumed, however with unrestricted $\Psi$ component, the estimate $\hat{\Psi}$ has approximately a %Toeplitz structure. Similarly, if we skip the diagonality assumption about $\Delta$, the non-diagonal %elements of its estimator are very close to zero.

%how to generalize
%Our model for the EEG/MEG could be extended to a sum of two, or more Kronecker products of three %components, similarly as suggested in \cite{Bijma2003} Each ingredient in the sum would represent %covariances within a particular type of activity. Moreover, the analysis could be done in a frequency domain %and each Kronecker product would correspond to one frequency band \cite{Bijma2008} and, in the context %of EEG/fMRI experiments, regressors could be obtained that are related to the amount of activity within each %frequency band.
In this paper we have given one illustration that shows the benefit of our method in the context of EEG/fMRI data. In a future study we will investigate the improvement of dipole localization in evoked response recordings when this new covariance framework is used instead of a covariance model that does not include the epoch/trial component $\Delta$. Yet another application of our method is the possibility of doing connectivity analysis on sensor level based on spontaneous EEG/MEG data using the estimated spatial covariance component $\hat{\Gamma}$. Although existing covariance models already yield estimated spatial components, the $\hat{\Gamma}$ that is output by our method is more reliable since it is corrected for the temporal and epoch/trial covariance patterns.

In sum, we have shown that our new covariance framework is a useful tool for analysis of EEG/MEG data in different contexts. It allows estimating a positive definite covariance matrix if the data of only one subject are available. Moreover, from the simulation studies we can conclude that the covariance estimate obtained with our iterative algorithm based on model \eqref{covariance_model} is very accurate if model \eqref{covariance_model} is plausible for the data. Our method can be used to analyze different EEG/MEG data sets. In the context of EEG-fMRI, the regressor extracted from EEG can be used as input for an integrated EEG-fMRI model, to model EEG variation independently of the choice of a specific frequency band or electrode site.

\appendix
\renewcommand{\thesubsection}{\Alph{section}\arabic{subsection}}
\section{Estimating a Toeplitz covariance matrix}
\label{Toeplitz_estimation}
\subsection{Notation}
\label{notation}
\noindent
We explain here how the problem of estimating a Toeplitz covariance matrix in our setting, where there are two additional components of the covariance matrix, is related to the problem in which a Toeplitz covariance matrix is estimated under the model without additional components. Let $\Gamma^{-1/2}$ and $\Delta^{-1/2}$ be positive definite square roots of $\Gamma^{-1}$ and $\Delta^{-1}$ respectively. 
If $Y_{1},\ldots,Y_{n}$ are defined as in \eqref{Toeplitz_equation}, then the columns of $\left[\left(\Delta^{-1/2}\otimes\Gamma^{-1/2}\right)Y_{1},\ldots,\left(\Delta^{-1/2}\otimes\Gamma^{-1/2}\right)Y_{n}\right]$ are independent and have a $\mathcal{N}\left(0,\Psi\right)$ distribution. Because the true covariance components $\Gamma,\Delta$ are not known, we use their current estimates and define $\tilde{Y}_{k}=\left(\hat{\Delta}^{-1/2}\otimes\hat{\Gamma}^{-1/2}\right)Y_{k}$ for $k=1,\ldots,n$. Given $\hat{\Gamma}$ and $\hat{\Delta}$, the columns of $\left[\tilde{Y}_{1},\ldots,\tilde{Y}_{n}\right]$ can be treated as independent vectors having a $\mathcal{N}\left(0,\Psi\right)$ distribution. Therefore they can be used in the estimation of the Toeplitz covariance matrix $\Psi$. There are $prn$ columns in total, therefore the effective sample size in the estimation of $\Psi$ equals $\tilde{n}=prn$. Finally, we denote the columns of $\tilde{Y}_{1},\ldots,\tilde{Y}_{n}$ by $x_{1},\ldots,x_{\tilde{n}}$.

\subsection{Maximum likelihood equation}
\label{derivation_Toeplitz_equation}
\noindent
If $x_{1},\ldots,x_{\tilde{n}}$ are independent realizations of $\mathcal{N}\left(0,\Psi\right)$ and $\Psi$ is a positive definite Toeplitz matrix defined by its first row $\left[\psi_{1},\ldots,\psi_{q}\right]$, the likelihood function equals
\begin{align*}
L\left(\Psi|x_{1},\ldots,x_{\tilde{n}}\right) =
\frac{1}{\sqrt{(2\pi)^{\tilde{n}q}}}\left|\Psi\right|^{-\frac{1}{2}\tilde{n}}
\text{etr}\left(-\frac{1}{2}\Psi^{-1}\sum_{k=1}^{\tilde{n}}{x_{k}x_{k}^{T}}
\right)
\end{align*}
and the maximum likelihood estimator of $\Psi$ must satisfy the property
\begin{align*}
\frac{\partial L\left(\Psi|x_{1},\ldots,x_{\tilde{n}}\right)}{\partial \psi_{i}}=0,
\end{align*}
for $i=1,\ldots,q$. It holds that 
\begin{align}
\label{dL_dpsi}
\frac{\partial L}{\partial\psi_{i}}=\left(\text{vec}\left(\frac{\partial L}{\partial\Psi}\right)\right)^{T}\left(\text{vec}\left(\frac{\partial\Psi}{\partial\psi_{i}}\right)\right),
\end{align}
and for calculating $\frac{\partial L}{\partial\Psi}$ and $\frac{\partial\Psi}{\partial\psi_{i}}$, the following equalities will be used, \cite{Magnus1999}:
\begin{align*}
\frac{\partial\left|\Psi\right|}{\partial \Psi}=\left|\Psi\right|\left(\Psi^{T}\right)^{-1};
\frac{\partial\text{tr}\left(\Psi^{-1}A\right)}{\partial \Psi}=-\left(\Psi^{-1}A\Psi^{-1}\right)^{T},
\end{align*}
for any $q\times q$ matrix $A$.
Using the first equality, the chain rule, and the symmetry of $\Psi$, it follows that
\begin{align*}
\frac{\partial}{\partial \Psi}\left|\Psi\right|^{-\frac{1}{2}\tilde{n}}=-\frac{1}{2}\tilde{n}\left|\Psi\right|^{-\frac{1}{2}\tilde{n}}\Psi^{-1}.
\end{align*}
If $A=\sum_{k=1}^{\tilde{n}}{x_{k}x_{k}^{T}}$, then from the second equality and the chain rule, we have that  
\begin{align*}
\frac{\partial}{\partial \Psi}\text{etr}\left(-\frac{1}{2}\Psi^{-1}A\right)=\frac{1}{2}\text{etr}\left(-\frac{1}{2}\Psi^{-1}A\right)\Psi^{-1}A\Psi^{-1}.
\end{align*}
Therefore 
\begin{align}
\frac{\partial L}{\partial\Psi}&=\left(2\pi\right)^{\frac{-q\tilde{n}}{2}}\left(\frac{\partial\left|\Psi\right|^{-\frac{1}{2}\tilde{n}}}{\partial\Psi}\text{etr}\left(-\frac{1}{2}\Psi^{-1}A\right)+\left|\Psi\right|^{-\frac{1}{2}\tilde{n}}\frac{\partial}{\partial\Psi}\text{etr}\left(-\frac{1}{2}\Psi^{-1}A\right)\right) \nonumber \\
&=\left(2\pi\right)^{-\frac{q\tilde{n}}{2}}\left(-\frac{1}{2}\tilde{n}\left|\Psi\right|^{-\frac{1}{2}\tilde{n}}\Psi^{-1}\text{etr}\left(-\frac{1}{2}\Psi^{-1}A\right)+\left|\Psi\right|^{-\frac{1}{2}\tilde{n}}\frac{1}{2}\text{etr}\left(-\frac{1}{2}\Psi^{-1}A\right)\Psi^{-1}A\Psi^{-1}\right) \nonumber \\
&=\frac{1}{2}\left(2\pi\right)^{-\frac{q\tilde{n}}{2}}\left|\Psi\right|^{-\frac{1}{2}\tilde{n}}\text{etr}\left(-\frac{1}{2}\Psi^{-1}A\right)\left(-\tilde{n}\Psi^{-1}+\Psi^{-1}A\Psi^{-1}\right). \label{dL_dPsi} 
\end{align}
Because $\Psi$ is symmetric and has a Toeplitz structure, it holds that 
\begin{align}
\label{dPsi_dpsi}
\frac{\partial \Psi}{\partial \psi_{i}}=V_{i},
\end{align}
where $V_{i}$ has ones on $i$th subdiagonals and zeros elsewhere.
Therefore, from \eqref{dL_dpsi}, \eqref{dL_dPsi} and \eqref{dPsi_dpsi}, we see that the likelihood equations $\frac{\partial L}{\partial\psi_{i}}=0$, $i=1,\ldots,q$ are
\begin{align*}
\frac{1}{2}\left(2\pi\right)^{-\frac{q\tilde{n}}{2}}\left|\Psi\right|^{-\frac{1}{2}\tilde{n}}\text{etr}\left(-\frac{1}{2}\Psi^{-1}A \right)
\left(\text{vec}\left(-\tilde{n}\Psi^{-1}+\Psi^{-1}A\Psi^{-1}\right)\right)^{T}\text{vec}\left(V_{i}\right)=0.
\end{align*}
They are satisfied if and only if
\begin{align*}
\text{vec}\left(-\tilde{n}\Psi^{-1}+\Psi^{-1}A\Psi^{-1}\right)^{T}\text{vec}\left(V_{i}\right)=0,
\end{align*}
which can be expressed as 
\begin{align*}
G\left(-\tilde{n}\Psi^{-1}+\Psi^{-1}A\Psi^{-1}\right)=\left(0,\ldots,0\right),
\end{align*}
which is the same as \eqref{Toeplitz_equation}, if one takes into account the definition of $A$, the relationship between $x_{1},\ldots,x_{\tilde{n}}$ and $Y_{1},\ldots,Y_{n}$, and the definition of $G$ given below \eqref{eq_for_Delta}.

\subsection{Estimating a Toeplitz covariance matrix}
\label{toeplitz_estimation}
\noindent
To estimate $\Psi$ by maximum likelihood, one can use the fact that a circulant positive definite covariance matrix can be estimated directly by maximum likelihood. If $C$ is an $l\times l$ positive definite circulant matrix, it can be parameterized as
\begin{align*}
C=
\begin{bmatrix}
c_{0}     & c_{l-1} & \dots  & c_{2} & c_{1}  \\
c_{1} & c_{0}    & c_{l-1} &         & c_{2}  \\
\vdots  & c_{1}& c_{0}    & \ddots  & \vdots   \\
c_{l-2}  &        & \ddots & \ddots  & c_{l-1}   \\
c_{l-1}  & c_{l-2} & \dots  & c_{1} & c_{0} \\
\end{bmatrix}
,
\end{align*}
with $c_{m}=c_{l-m}$, for $m=1,\ldots,l-1$.

If $z_{1},\ldots,z_{\tilde{n}}$ are independent random vectors having a $\mathcal{N}\left(0,C\right)$ distribution and $\hat{S}=\frac{1}{\tilde{n}}\sum_{k=1}^{\tilde{n}}{z_{k}z_{k}^{T}}$ is the sample covariance matrix, the maximum likelihood estimators of the entries of the circulant covariance matrix equal
\begin{equation}
 \hat{c}_{u}=\frac{1}{l}\sum_{(i-j)\equiv u\text{mod}l}{\hat{S}_{i,j}},
\label{circulant_equation}
\end{equation}
for $u=0,\ldots,l-1$, \cite{Olkin1969}. This means that $\hat{c}_{u}$ is the average of entries of the sample covariance matrix that correspond to positions of $c_{u}$ in $C$.
If $l\geq q$, then the $q\times q$ upper left block of $C$ is a positive definite Toeplitz matrix. If $l\geq 2q-1$ is sufficiently large, then for any given $\Psi$ that is Toeplitz and positive definite, there exists an $l\times l$ circulant non-negative definite covariance matrix in which $\Psi$ can be embedded, \cite{Dembo1989}. The idea is to estimate $C$ and the upper left block of this estimate will be the estimate of the Toeplitz covariance matrix $\Psi$.

The problem is that estimating $C$ requires vectors $z_{1},\ldots,z_{\tilde{n}}$ of length $l$, while we only have data consisting of vectors $x_{1},\ldots,x_{\tilde{n}}$ of length $q$. However, one can use the expectation maximization algorithm \cite{Dempster1977}, which uses the available data and the current estimate of the circulant covariance matrix $\widetilde{C}$ to replace the missing data by their expectations conditioned on $x_{1},\ldots,x_{\tilde{n}}$ and $\widetilde{C}$. Then these data are used to update the estimate of the circulant component. This alternating procedure is repeated until convergence. The upper left block of the final estimate of $C$ is our numerical approximation of the maximum likelihood estimate of the Toeplitz covariance matrix $\Psi$.

Estimates of $\Gamma,\Psi,\Delta$ are updated iteratively, using Algorithm~\ref{alg:algorithm1}, in which updating an estimate of $\Psi$ given current estimates of $\Gamma$ and $\Delta$ is done in an inner loop. Let $m$ be the iteration index of the outer loop of Algorithm~\ref{alg:algorithm1}. 
Let $\tilde{\Psi}^{(0)}$ be the initial value for the Toeplitz covariance matrix with its circulant extension $\tilde{C}^{(0)}$. If $m=1$, we take $\tilde{C}^{(0)}$ equal to the identity matrix, otherwise we take the initial value from the iteration $m-1$ of Algorithm~\ref{alg:algorithm1}: $\tilde{C}^{(0)}=\hat{C}_{m-1}$. The $i$th step of the estimation procedure for $\hat{\Psi}_{m}$ consists of the following steps:
\begin{enumerate}
\item Define $\tilde{A}^{(i)},\tilde{P}^{(i)},\tilde{U}^{(i)}$ by
\begin{align*}
\left[\tilde{C}^{(i)}\right]^{-1}=
\begin{bmatrix}
\tilde{A}^{(i)} & \tilde{P}^{(i)T} \\
\tilde{P}^{(i)} & \left(\tilde{U}^{(i)}\right)^{-1}
\end{bmatrix},
\end{align*}
where $\tilde{P}^{(i)}$ is an $\left(l-q\right)\times q$ matrix, $\tilde{U}^{(i)}$ is a $\left(l-q\right)\times\left(l-q\right)$ matrix.
\item E-step: compute estimate of data
\begin{align*}
\hat{z}_{k}^{(i)}=
\begin{bmatrix}
x_{k} \\
-\tilde{U}^{(i)}\tilde{P}^{(i)}x_{k} 
\end{bmatrix}
\end{align*}
for $k=1,\ldots,\tilde{n}$.
\item Compute
\begin{align*}
\hat{S}^{(i)}=\frac{1}{\tilde{n}}\sum_{k=1}^{\tilde{n}}{\hat{z}_{k}^{(i)}\hat{z}_{k}^{(i)T}}+
\begin{bmatrix}
0 & 0 \\
0 & \tilde{P}^{(i)}
\end{bmatrix}
\end{align*}
\item M-step: maximize the likelihood with respect to the circulant covariance matrix using $\hat{S}^{(i)}$, by formula \eqref{circulant_equation}. Denote the maximizer by $\tilde{C}^{(i+1)}$.
\end{enumerate}
These steps are repeated until convergence. An approximation of the estimate of the Toeplitz covariance matrix $\Psi$ is the $q\times q$ upper left block $\tilde{\Psi}$ of the final estimate of the circulant covariance matrix $\tilde{C}$. If $\hat{\Gamma}_{m-1}$ and $\hat{\Delta}_{m-1}$ are the current estimates of $\Gamma$ and $\Delta$, the update of the Toeplitz component will be $\hat{\Psi}_{m}=\tilde{\Psi}$ with its circulant extension $\hat{C}_{m}=\tilde{C}$.

\cite{Dembo1989} and \cite{Newsam1994} give conditions for $l$ that guarantee convergence to a maximum likelihood estimator of a Toeplitz covariance matrix, but they are impractical in our context. Instead we use $l=2q-1$, which is feasible in practice. Moreover, we conclude from the simulation studies that our estimation procedure provides accurate estimates of $\Psi$.

\section*{Acknowledgements}
\noindent
This work was financially supported by a STAR-cluster grant from the Netherlands Organization of Scientific Research.

%\bibliographystyle{apalike}
%\bibliography{science}

\end{document}